\documentclass[aps,prb,superscriptaddress,twocolumn,floatfix,footinbib]{revtex4-1}
\usepackage[utf8]{inputenc}

\usepackage{graphicx}
\usepackage[colorlinks = true, linkcolor = blue, urlcolor  = blue, citecolor = blue, anchorcolor = blue]{hyperref}

\usepackage{multirow}
\usepackage{makecell}
\usepackage{amsmath}
\usepackage{upgreek}

\newcommand{\geff}{$g_\mathrm{eff}$}

\newcommand {\Sec}[1] {Section~\ref{#1}}
\newcommand {\Eq}[1] {Eq.~\ref{#1}}

\newcommand {\Fig}[1] {Figure~\ref{#1}}
\newcommand {\Tab}[1] {Table~\ref{#1}} 

\newcommand{\beq}{\begin{equation}}
\newcommand{\eeq}{\end{equation}}

\newcommand{\microsec}{$\upmu$s}

\newcommand{\beqa}{\begin{eqnarray}}
\newcommand{\eeqa}{\end{eqnarray}}

\newcommand{\ttwo}{$T_2$}
\newcommand{\ttwostar}{$T_2^*$}

\newcommand{\tone}{$T_1$}
\newcommand{\tonen}{$T_{\mathrm{1n}}$}
\newcommand{\tonee}{$T_{\mathrm{1e}}$}

\newcommand{\ttwon}{$T_{\mathrm{2n}}$}
\newcommand{\ttwoe}{$T_{\mathrm{2e}}$}

\usepackage{bm}
\newcommand{\angstrom}{\textup{\AA}}

\usepackage{soul}

\begin{document}
	\title{Coherent spin dynamics of ytterbium ions in yttrium orthosilicate}
	
	\author{Hee-Jin Lim}
	\affiliation{London Centre for Nanotechnology, University College London, London WC1H 0AH, UK}
	\author{Sacha Welinski}
	\affiliation{PSL Research University, Chimie ParisTech, CNRS, Institut de Recherche de Chimie Paris, 75005, Paris, France}
	\author{Alban Ferrier}
	\affiliation{PSL Research University, Chimie ParisTech, CNRS, Institut de Recherche de Chimie Paris, 75005, Paris, France}
	\affiliation{Sorbonne Universit\'es, UPMC Univ Paris 06, 75005, Paris, France}
	\author{Philippe Goldner}
	\email{philippe.goldner@chimie-paristech.fr}
	\affiliation{PSL Research University, Chimie ParisTech, CNRS, Institut de Recherche de Chimie Paris, 75005, Paris, France}
	\author{J.J.L. Morton}
	\email{jjl.morton@ucl.ac.uk}
	\affiliation{London Centre for Nanotechnology, University College London, London WC1H 0AH, UK}
	\affiliation{Dept.\ of Electronic \& Electrical Engineering, University College London, London WC1E 7JE, UK}
	
	\date{\today}

	\begin{abstract} 
	We investigate the electron and nuclear spin coherence properties of  ytterbium ($\mathrm{Yb}^{3+}$) ions with non-zero nuclear spin, within an yttrium orthosilicate (Y$_2$SiO$_5$) crystal, with a view to their potential application in quantum memories or repeaters. 
	We find electron spin-lattice relaxation times are maximised at low magnetic field ($<100$~mT) where $g~\sim6$, reaching 5~s at 2.5~K, while coherence times are maximised when addressing ESR transitions at higher fields where  $g\sim0.7$ where a Hahn echo measurement yields \ttwo\ up to 73 $\upmu$s. 
	Dynamical decoupling (XY16) can be used to suppress spectral diffusion and extend the coherence lifetime to over 0.5 ms, close to the limit of instantaneous diffusion.
	Using Davies electron-nuclear-double-resonance (ENDOR), we performed coherent control of the $^{173}\mathrm{Yb}^{3+}$ nuclear spin and studied  its relaxation dynamics. At around 4.5~K we measure a nuclear spin \tone\ and \ttwo\ of 4 and 0.35 ms, respectively, about 4 and 14 times longer than the corresponding times for the electron spin. 
	\end{abstract}
 
\maketitle

\section*{Introduction}

Paramagnetic rare earth (RE) ions in optical crystals are rich systems possessing electron and nuclear spins, and optical transitions,\cite{Goldner:2015} making them attractive for coherent interactions with both optical and microwave photons\cite{Afzelius:2013,Probst:2013vp,Williamson:2014fb,Wolfowicz:2015exa,Rancic:2017bn}. 
The excellent coherence properties of RE optical transitions have lent themselves to photon memories for quantum repeaters\cite{Tittel:2010bp,Bussieres:2014dc,Saglamyurek:2015esa}, while their nuclear spin degree of freedom has demonstrated the capability for long-term coherent storage of quantum information\cite{Zhong:2015bw,Wolfowicz:2015exa,Jobez:2015gt,Rancic:2017bn,Kutluer:2017ci}.
Embedded in microwave cavities, the collective spin dynamics of ensembles of paramagnetic RE ions have been investigated with a view to develop efficient and faithful microwave memories\cite{Probst:2013vp,Chen:2016uy} and microwave-optical conversion.\cite{Williamson:2014fb,OBrien:2014dc}

Compared to other paramagnetic RE ions used for quantum memories, like $\mathrm{Er}^{3+}$ [\onlinecite{Saglamyurek:2015esa}, \onlinecite{Dajczgewand:2014et}] and $\mathrm{Nd}^{3+}$ [\onlinecite{Zhong:2017fe}, \onlinecite{Bussieres:2014dc}], ytterbium ions ($\mathrm{Yb}^{3+}$) have a number of potential advantages. \cite{Kis:2014fq,Bottger:2016ix,Welinski:2016gi} As with $\mathrm{Er}^{3+}$and $\mathrm{Nd}^{3+}$, the optical transitions of ytterbium are accessible via single mode laser diodes in the near infrared.  
However, the lower nuclear spin quantum number of ytterbium,
(e.g.~$I=1/2$ of $^{171}\mathrm{Yb}^{3+}$ and $I=5/2$ of $^{173}\mathrm{Yb}^{3+}$), with respect to the $^{145}\mathrm{Nd}^{3+}$ and $^{167}\mathrm{Er}^{3+}$ (both $I=7/2$), is beneficial for addressing optical transitions between ground and excited spin states and initialisation into a ground spin state. Coherence times up to 130 $\mu$s have been also observed for electron spin resonance transitions in $\mathrm{Yb}^{3+}\mathrm{:CaWO}_4$.\cite{Rakhmatullin:2009jfa}

Single crystal yttrium orthosilicate  ($\mathrm{Y}_2\mathrm{SiO}_5$, commonly referred to as YSO) has been a reference host material for quantum information processing research 
and applications, mainly due to 
its low natural abundance of nuclear spins which would otherwise lead to spin decoherence.\cite{Bottger:2006jo,Arcangeli:2014dr}
$\mathrm{Yb}^{3+}$ in YSO has already been shown to exhibit 
good optical properties like high oscillator strengths, low inhomogeneous linewidths and favourable branching ratios into narrow transitions in comparison to
$\mathrm{Er}^{3+}$, $\mathrm{Pr}^{3+}$, and $\mathrm{Eu}^{3+}$ in the same host\cite{Welinski:2016gi}.
The maximal electron g-factor ($g_\mathrm{max}\sim6$) of the Yb:YSO ground state lies somewhere between that of other paramagnetic REs such as Nd:YSO ($g_\mathrm{max}\sim 4.2$)\cite{Wolfowicz:2015exa} and Er:YSO ($g_\mathrm{max}\sim 15.5$)\cite{2006PhRvB..74u4409G}. A larger g-factor is beneficial in enhancing the cooperativity 
between spins and microwave cavities, important in developing microwave quantum memories\cite{Probst:2013vp} or microwave-to-optical quantum transducers\cite{FG:2015dr}. However, larger g-factors often come at expense of increased decoherence rates, due to stronger coupling to other spins, leading to spectral diffusion and instantaneous diffusion.\cite{Rakhmatullin:2009jfa}
 A detailed understanding of decoherence and relaxation processes for paramagnetic RE ions is therefore essential for identifying the optimum species, sites and transitions to address for the different quantum technological applications. 

Here, we investigate the relaxation and the decoherence dynamics of electron and nuclear spins of Yb:YSO, using pulsed electron spin resonance (ESR) at X-band (9.8~GHz), and electron nuclear double resonance (ENDOR), in the temperature range 2--8 K. We identify multiple electron spin-lattice relaxation processes in Yb:YSO, combining our pulsed ESR measurements with optical spectroscopy of the ground state multiplet\cite{Welinski:2016gi}. We find that spectral diffusion\cite{Mims:1968dg,Rakhmatullin:2009jfa} is the most common source of decoherence for electron spins, further evidenced by stimulated echo decay measurements, and are able to suppress it considerably using XY16 dynamical decoupling~\cite{Viola:1998ky}.
Using Davies ENDOR we explored the coherent spin properties of the $^{173}\mathrm{Yb}$ nuclear spins, including nuclear spin Rabi oscillations, studying inhomogeneous broadening through the nuclear spin \ttwostar, nuclear spin relaxation time \tonen, and the nuclear spin coherence time \ttwon.

\begin{figure*}
 	\centerline{\includegraphics[width=18cm]{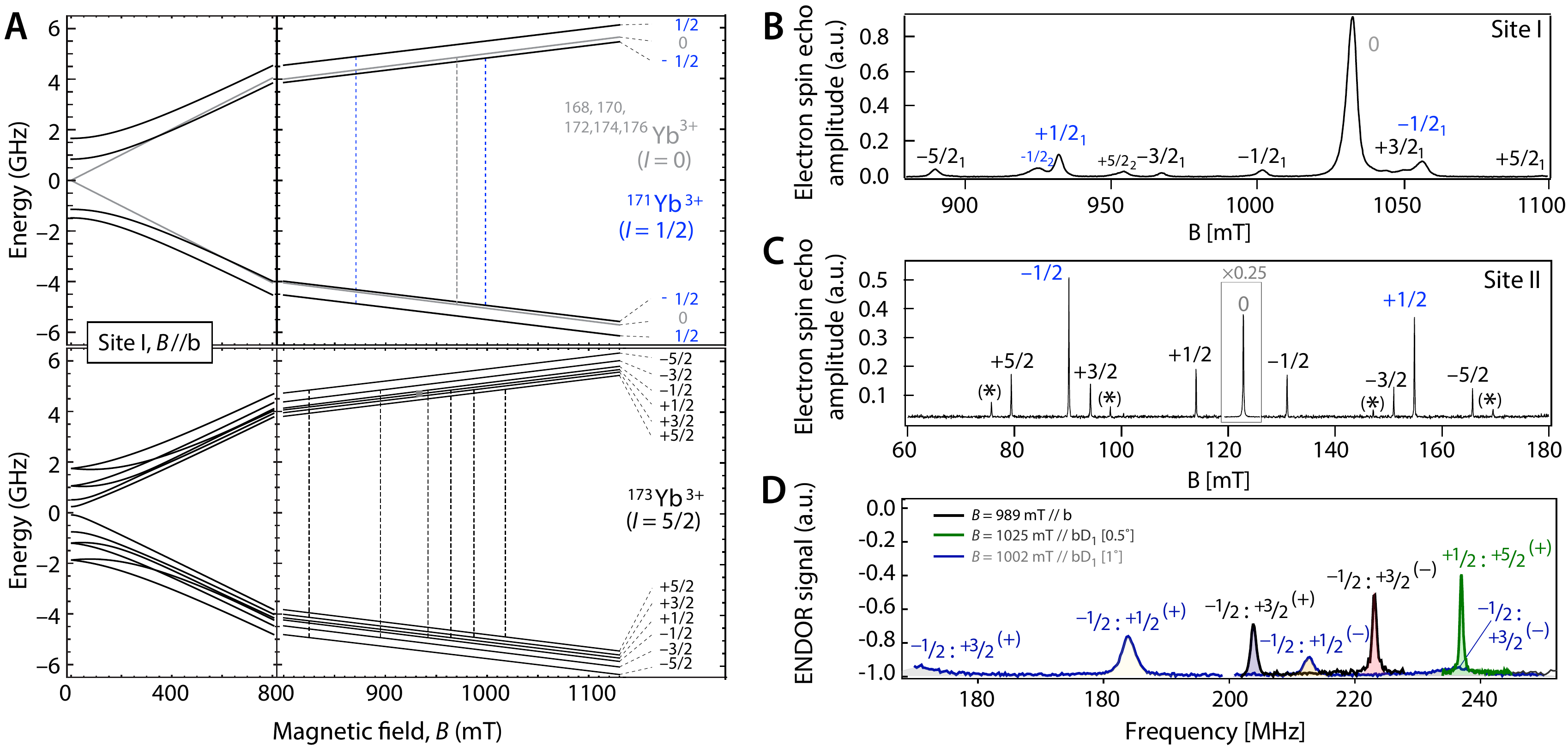}}
 	\caption{ 
	\textbf{(a)} Spin energy level diagram for the various isotopes of Yb, in Site I, as a function of magnetic field $B$ applied along the crystal axis $b$. Dashed lines indicate allowed ESR transitions ($\Delta m_I=0$) at 9.8~GHz, corresponding to peaks observed in		
	\textbf{(b)} the corresponding electron spin echo-detected field sweep (EDFS) spectrum measured at 7~K.  A similar EDFS spectrum for site II is shown in panel 
	\textbf{(c)}. In both spectra, peaks are labelled according to nominal $m_I$ values for convenience, although the spin states show considerable mixing. 
	In addition, peaks labelled with ($\star$) correspond to forbidden ESR transitions ($\Delta m_I = \pm 1$).	
	\textbf{(d)} Davies electron nuclear double resonance (ENDOR) spectra of $^{173}\mathrm{Yb}$ in site I, at 5~K. Peaks are labeled according to our best estimate of the $m_I$  states involved, with the superscript indicating the upper ($^+$) or lower ($^-$) electron spin manifold.}
	\label{fig:spectra}
 \end{figure*}

\section{Electron and nuclear spin spectroscopy} \label{sec:spectra}

The sample studied is a Czochralski-grown YSO crystal doped with $\mathrm{Yb}^{3+}$ (natural isotopic abundance), at a nominal concentration of 0.005 at.~\% (50 ppm). 
Yb isotopes with non-zero nuclear spin number are $^{171}\mathrm{Yb}$ (\textit{I}=1/2) and $^{173}$Yb (\textit{I}=5/2), which respectively constitute 14\% and 16\% of the total Yb concentration, with the remaining 70\% comprised of $I=0$ isotopes.
In YSO, which has a monoclinic structure and $C_{2h}^6$ (C2/c) space group, Yb can substitute Y located in two crystallographic sites with $C_1$ point symmetry\cite{Wen:2014eh}, denoted site I and site II\cite{Kurkin:1980jx,Welinski:2016gi}.
Each site has two sub-sites which are magnetically equivalent only when the applied magnetic field is  parallel or perpendicular to the $C_2$ symmetry axis (the crystal axis $b$). The g-factor ($g$) and hyperfine ($A$) tensors for Yb in each of these sites have been extracted from earlier continuous wave (CW) ESR measurements\cite{Welinski:2016gi}, yielding ESR transitions with a wide range of effective g-factors ($g_{\mathrm{eff}}$) from 0.5 to 6, and hyperfine coupling strengths of up to 2.5~GHz (see Fig.S2\cite{supp}). 

To ensure consistency with the earlier CW ESR studies, we first show electron spin echo-detected field sweep (EDFS) obtained using the two-pulse echo sequence ($\pi/2$-$\tau$-$\pi$-$\tau$-echo) with $\tau =1~\upmu$s, microwave pulse durations of 16 and 32~ns, and a 250~ns integration window. 
While this method is suitable for determining the spectral position of each transition, interpretation of the peak height requires knowledge of the transition dipole strength ($\gamma_x^e$) which varies considerably as a function of magnetic field and across different sites (Eq.S3\cite{supp}).

\Fig{fig:spectra}(a) illustrates the allowed ESR transitions expected for Site I, with the magnetic field $B_0$ applied close the crystal axis $b$ and \geff\ is 0.70. Due to the imperfect alignment with $b$ (estimated to be of order 2$^{\circ}$), the spectrum of only one sub-site is seen in each plot. At X-band, the spin eigenstates are considerably mixed so $\Delta m_{S,I}$ are not good quantum numbers and are only used as a qualitative identification of the states.
\Fig{fig:spectra}(b,c) shows the echo-detected ESR spectra ($B_0\parallel b$) from Yb ions in site I and II, highlighting the large difference in \geff\ between the two sites. For each site, sets of ESR peaks can be identified from their hyperfine coupling to the $^{171}\mathrm{Yb}$ and $^{173}\mathrm{Yb}$ isotopes, in addition to a single, intense resonance from the family of Yb isotopes with zero nuclear spin. The peaks have  intensities  consistent with the natural isotopic composition of Yb and arise primarily from allowed ESR transitions ($\Delta m_I = 0$), though a small number of forbidden transitions ($\Delta m_I = \pm 1$) are also weakly visible. We extract linewidths of the $m_I=1/2$, $^{171}\mathrm{Yb}$ resonance for site I and site II of 3.4~mT and 0.1~mT respectively, which can be compared to their respective $g_\mathrm{eff}$ of 0.7 and 6. These linewidths are consistent with previous CW ESR measurements\cite{Welinski:2016gi} and attributed to a site-dependent g-strain ($\Delta g/g \sim$ 0.1--0.3$\%$) similar to that observed in $\mathrm{Er}^{3+}$:YSO with the same ion concentration\cite{Welinski:2017ew}.

We next turn to the nuclear spin transitions, studied for the $^{173}\mathrm{Yb}$ isotope in site I using the Davies ENDOR technique\cite{Davies:1974gu,schweiger2001principles} with a Tidy pulse\cite{MOSiBiENDOR,Tyryshkin2006}, RF $\pi$-pulse duration of 1.5~\microsec\, and the magnetic field aligned approximately with $b$ (see \Fig{fig:spectra}(d)). 
ENDOR spectra are shown measured using three different allowed ESR transitions (corresponding to $m_I=\{-1/2, +3/2,+5/2$\}), each containing ENDOR peaks attributed to nominally allowed ($|\Delta m_I|=1$) and forbidden ($|\Delta m_I|=2$) nuclear spin transitions in the upper($^+$) and lower($^-$) electron spin manifolds.
Measured ENDOR linewidths vary from 0.6 to 2.6~MHz, which we attribute respectively to the excitation bandwidth of the RF pulse and A-strain ($\Delta A/A \leq 1\%$). 
The effect of A-strain is visible as the sensitivity of the ENDOR transition frequency to $A$ ($df/dA$) is a strong function of crystal orientation, leading to large changes in ENDOR linewidth under sample rotations of only 1$^\circ$ (see Fig.S1a and Fig.S2c)\cite{supp}.
 
 \section{Electron spin-lattice relaxation}\label{sec:SLR}

  \begin{figure}
  	\centerline{\includegraphics[width=7.5cm]{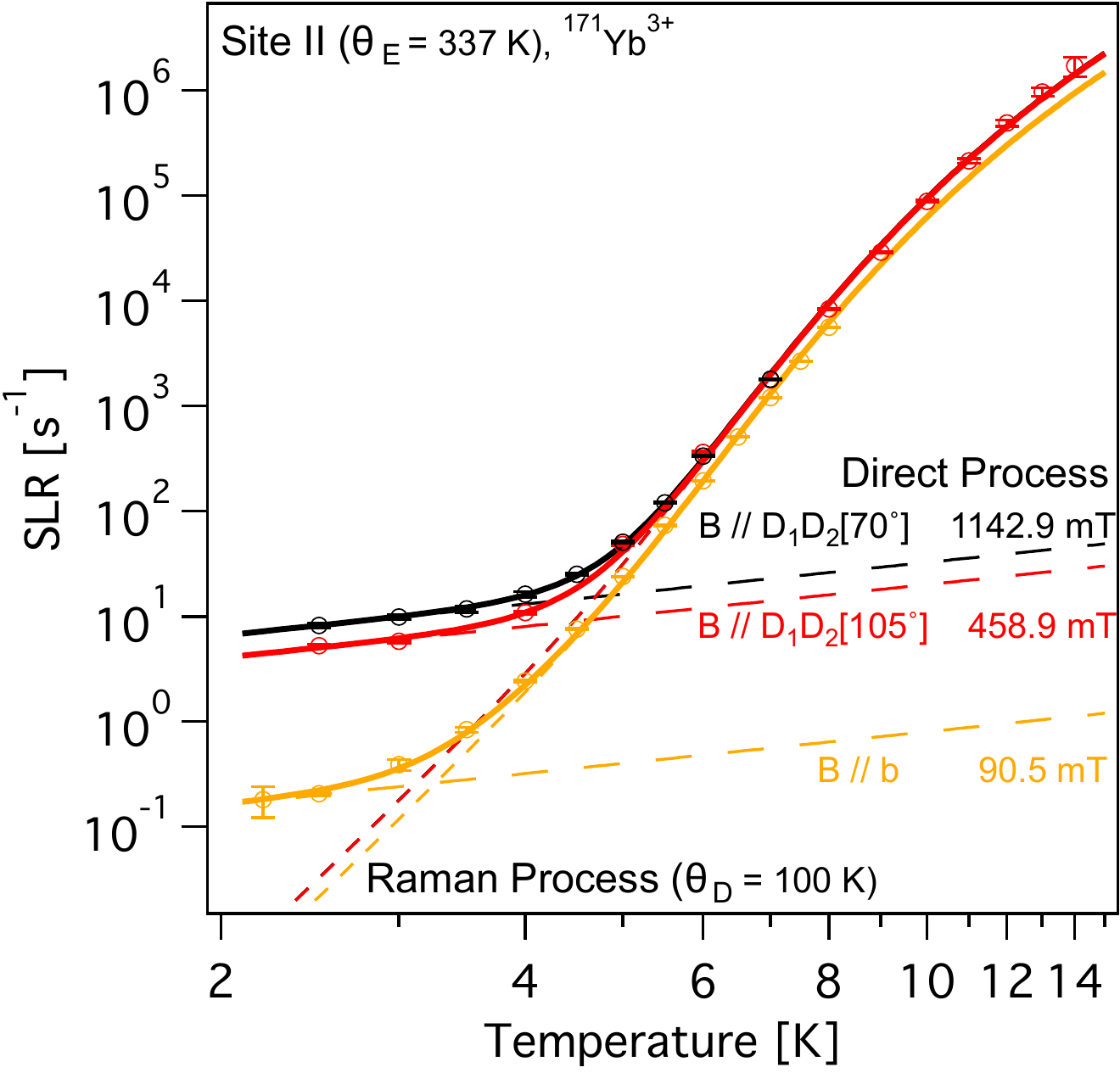}}
  	\caption{Electron spin relaxation rates of Yb:YSO in site II, measured using an inversion recovery sequence. Solid lines are fits based on a model comprising both one-phonon (dashed line) and two-phonon (dotted line) processes. The values extracted from these fits are presented in \Tab{tab:SLR}.} \label{fig:SLR}
  \end{figure} 

Electron spin relaxation, commonly caused by spin-phonon coupling\cite{VanVleck:1940ji,Shrivastava:1983tr,abragam2012electron} and characterised by the timescale \tonee, impacts the coherence time of the electron spin (\ttwoe) and nuclear spin (\ttwon) in several ways. First, there is the `direct' impact on the central spin where \ttwoe\ is bounded by \tonee, and \ttwon\ bounded by 2\tonee (assuming a strong hyperfine coupling and in the regime where the thermal electron spin polarisation is much less than one\cite{MORTON2008315}). Second, there is an `indirect' impact where spin-flips of neighbouring electron spins lead to spectral diffusion of the central spin~\cite{Tyryshkin:2012fi}. 
We study \tonee\ of Yb ions in site I and II at various crystal orientations and in the temperature range 2--10~K, using the inversion-recovery method ($\pi-\tau_r-\pi/2-\tau_e-\pi-\tau_e-{\rm echo}$, where $\tau_r$ is swept), as shown in \Fig{fig:SLR}.
  
At low temperatures ($\lesssim4$~K), we find the electron spin-lattice relaxation rate is inversely proportional with temperature, which we attribute to a direct one-phonon process, which occurs via interaction with a phonon resonant with the spin transition\cite{Orbach:1961cd}. At higher temperatures, two-phonon relaxation processes dominate, which have a much stronger temperature dependence as we discuss further below.

The direct one-phonon processes is enabled by the mixing of crystal field levels by an applied magnetic field, and hence its rate $R_\mathrm{1p}$ has a strong magnetic field dependence being proportional to $g_\mathrm{eff}^2 B^4$ in the regime where the Zeeman splitting is much less than $k_BT$:\cite{Shrivastava:1983tr,PhysRevB.77.085124}.
 
\beq
R_\mathrm{1p} = \alpha_D (\hat \theta) g_\mathrm{eff}^3 B^5 \coth \left(\frac{\mu_B g_\mathrm{eff} B}{2 k_B T}\right) \approx  \frac{2\alpha_D (\hat \theta)k_B T g_\mathrm{eff}^2 B^4 }{\mu_B}.\label{eq:SLR1p}
\eeq 

The constant $\alpha_D$ varies weakly with site\cite{Kurkin:1980jx}, crystal orientation\cite{Mikkelson:1965kb} and its values under various conditions, extracted from fits to the plots in \Fig{fig:SLR}, are shown in \Tab{tab:SLR}. For example, we see that for the field orientation $D_1D_2[70^{\circ}]$ the crystal field mixing appears significantly reduced leading to a three-fold reduction in $\alpha_D$, and also that the crystal field mixing appears slightly larger in site I than in site II. Given a constant Zeeman energy splitting (e.g.~when considering transitions addressed by X-band ESR), $R_\mathrm{1p}$ has an effective $B^2$ dependence, which can be seen when comparing the low-temperature \tonee\ values for $^{171}$Yb site II at several orientations. 

	\begin{table}
	\centering
	\begin{tabular}{cc|c|cc|ccc} 
		\Xhline{3\arrayrulewidth}
		Site & \makecell{$\theta_E$ \\ {[}K{]}} &Isotope & \makecell{$B$ \\ {[}mT{]}} &Angle & \makecell{ $\alpha_D$ \\ {[}Hz/T$^5${]} }& \makecell{$\theta_D$ \\ {[}K{]}} & \makecell{$\alpha_R$ \\ {[}$\times 10^{18}$ \\ Hz $\cdot$K$^4${]}}\\
		\Xhline{3\arrayrulewidth}
		\multirow{2}{*}{I} & \multirow{2}{*}{160} &171 &1020.8 &$b$ &13.2 & \multirow{2}{*}{100} &0.88 \\ 
		& &173 &985.6 &$b$ &10.1 &  &0.54 \\ \hline
		\multirow{4}{*}{II} & \multirow{4}{*}{337}  & \multirow{4}{*}{171} &1142.9 &\makecell{D$_1$D$_2$[70$^\circ$]} & 1.7 & \multirow{4}{*}{100} &2.4 \\
		& & &798.4 &\makecell{D$_2$ \\ (D$_1$D$_2$[90$^\circ$]) } & 2.0 &  &2.4 \\ 
		& & &458.9 & D$_1$D$_2$[105$^\circ$] &6.5 & &2.4 \\ 
		& & &90.5 &$b$ &6.7 & 
		&1.6 \\ 
		\Xhline{3\arrayrulewidth}
	\end{tabular}
	\caption{Parameters to describe spin-lattice relaxation. $\alpha_D$ is for the direct process, and $\alpha_R$ for the two-phonon process. $\theta_D$ is a temperature corresponding to the maximum phonon energy contributing to the spin relaxation.}
	\label{tab:SLR}
\end{table}

Two-phonon spin relaxation\cite{abragam2012electron} has been categorised by resonant (Orbach) and non-resonant (Raman) processes\cite{Orbach:1961cd}. 
The Orbach process emerges when temperature is high enough to excite phonons resonant with some high-lying state (with energy $k_B\theta_E$ above the ground state) which mediates an emission and an absorption of a phonon, and has a temperature dependence of $e^{-\theta_E/T}$. The Raman process is similarly a two-phonon mechanism, but not resonant with a particular excited state and showing a $T^9$ temperature dependence. 
Fitting our experimental values to a combined model with these processes produces values for $\theta_E$ of 97 K (site I) and 107 K (site II) --- these are similar to those reported previously\cite{Kurkin:1980jx} but not consistent with the actual energies of the first excited $^2F_{7/2}$ state (160~K (site I) and 337~K (site II), as measured in the optical spectroscopy)\cite{Campos:2004cv, Welinski:2016gi}.

Consistent with previous studies on paramagnetic RE ions\cite{1967PhRv161.386K,Shrivastava:1983tr}, we therefore adopt a more general description of the two-phonon process, which takes account of the maximum phonon energy ($k_b \theta_D$) and the actual $\theta_E$ as measured (e.g.) by optical spectroscopy\cite{Welinski:2016gi}:
\begin{align}
& R_\mathrm{2p} (\alpha_R, \theta_D,T ; \theta_E) = \nonumber \\  &\quad \alpha_R(\hat \theta) \int_0^{\frac{\pi}{2}}\frac{q^8 e^{-\frac{\theta_D}{T} \sin q}  \mathrm d q}{\left(1-e^{-\frac{\theta_D}{T} \sin q}\right)^2 \left(\theta_E^2 - \theta_D^2 \sin^2 q\right)^2}
 \label{eq:SLR}
\end{align}
The assumption is that the low energy branch of phonons will be most effective at driving spin relaxation at low temperatures\cite{Senyshyn:2004hk}. Values for the free parameters $\alpha_R$ and $\theta_D$, which describe the general two-phonon process, were extracted from fits to the experimental results (see, e.g.~Fig.\ref{fig:SLR}) and presented in Table \ref{tab:SLR}. Full datasets for all transitions measured are shown in Sec.S3\cite{supp}.

The data suggest $\theta_D$ of about 100~K both for site I and site II, and as there is no reason to assume this value should be different for the two sites, we constrain this value to be constant for all datasets. The value of $\theta_D$ is consistent with Raman phonon spectra\cite{Campos:2004cv} which identify lowest optical phonon mode at $<$ 100 cm$^{-1}$ (140 K), implying the acoustic phonon cut-off is below 140 K.
While both sites have the similar $\theta_D$, the Raman process is slower for site II due to the large $\theta_E$.
We have not observed any notable difference in \tonee\ between Yb isotopes (see Fig.S3 and Fig.S4)\cite{supp}.

\section{Electron spin coherence time}\label{sec:EPRDD}
   
Having determined the bounds on spin coherence lifetimes from spin-lattice relaxation, we now turn to measurements of electron spin coherence through two-pulse (Hahn) echo measurements, and dynamical decoupling (DD) schemes such as XY16~\cite{Viola:2003fc,Khodjasteh:2009ff}.
We focus on $^{171}\mathrm{Yb}$ in site I with $B = 1020.8$~mT applied approximately parallel to the $b$ axis. In this orientation, $g_\mathrm{eff}$ is low (0.7) and electron spin echo envelope modulation (ESEEM) from $^{89}\mathrm{Y}$ nuclear spins is negligible due to the weak superhyperfine interaction. 
   
\begin{figure}
   	\centerline{\includegraphics[width=8.5cm]{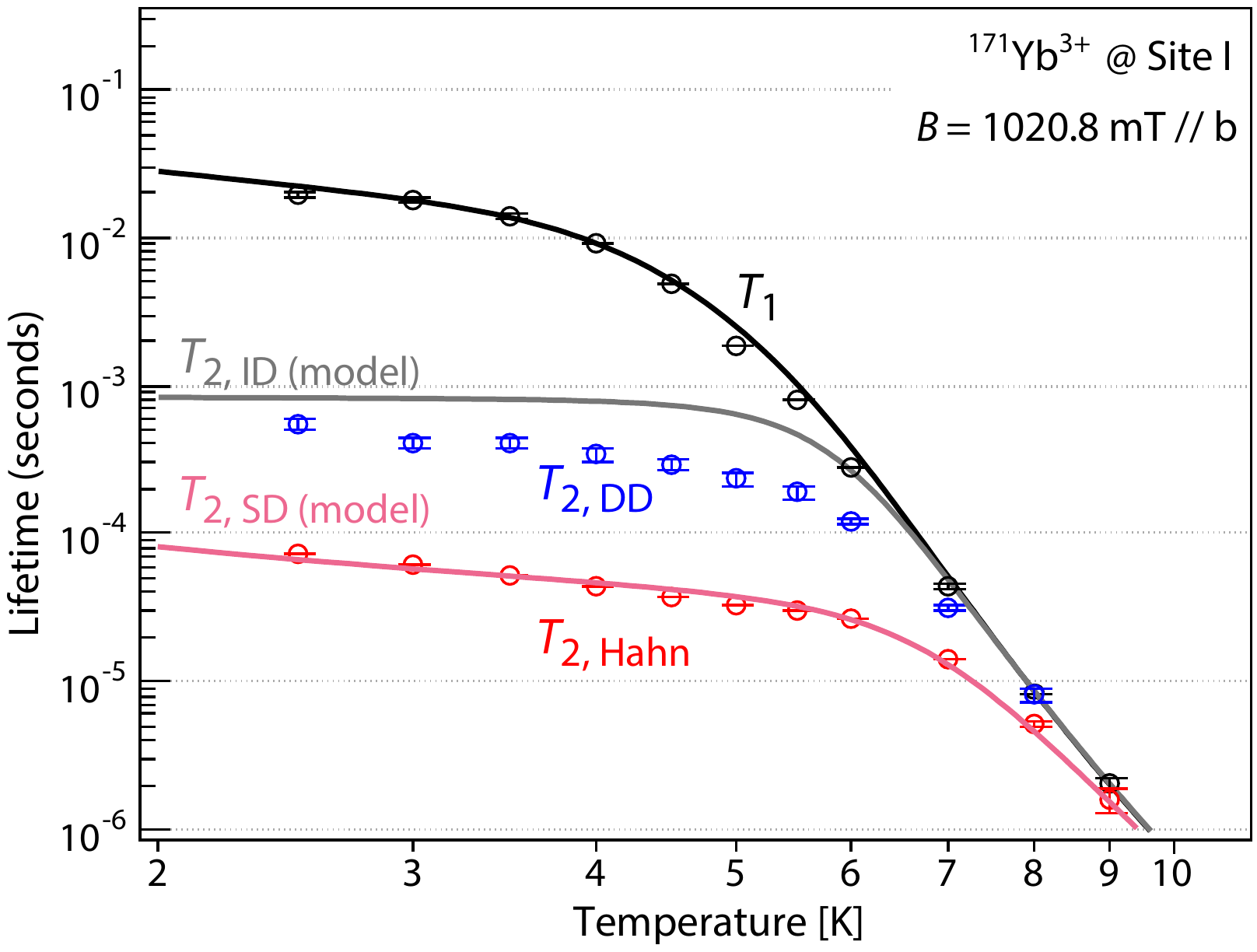}}
   	\caption{Coherence lifetime of electron spins of $^{171}\mathrm{Yb}^{3+}$(I=1/2) in site I at $B$ of 1020.8 mT applied along the crystal axis $b$.
   	The black circles are relaxation lifetimes ($T_1$) extrapolated from the inversion recovery measurements (a replica of the data shown in Fig.\ref{fig:SLR}a).
   	The coherence lifetimes obtained the two-pulse echo technique (Hahn) are shown as red dots, and denoted by $T_2$.
   	$T_\mathrm{DD}$ denotes coherence lifetimes under the XY16 dynamic decoupling scheme.
   	The grey curve is for the $T_2$ bound by the instantaneous diffusion (ID).
   	The red curve includes the SD caused by interactions between site I and site II.
	} \label{fig:eprddtem}
   \end{figure} 
   
We find $T_2$ is bounded by $T_1$  for temperatures above about 8~K, but at lower temperatures additional decoherence mechanisms are visible, as shown in \Fig{fig:eprddtem}. The longest measured value for  $T_{2,{\rm Hahn}}$ was 73 $\upmu$s, measured at 2.5~K.
Below 8~K, the electron spin decoherence  followed stretched exponential decay\cite{Hu:1974kx,Maryasov:1982cg,PhysRevB.77.085124} of the form $\exp \left[-(2\tau/T_2)^n \right]$, with the stretch parameter $n$ rising to 2.7 at 2.5~K, consistent with spectral diffusion\cite{Mims:1968dg}.
   
We first consider spectral diffusion experienced by a central spin in (e.g.)\ site I, arising \emph{only} from spin-flips of its neighbours in sites I and II (we neglect processes such as spin flip-flop terms due to the large inhomogeneous spin linewidth compared to average dipole coupling strength). 
According to a Lorentz diffusion model\cite{Klauder:1962it}, the contribution of spectral diffusion to the $T_2$ of this spin, can be expressed as~\cite{Mims:1968dg,Hu:1974kx,Bottger:2006jo,supp}:
\beq
T_\mathrm{SD, I}^{-1} = \frac{\pi}{6}\sqrt{\frac{\mu_0hn}{\sqrt{3}}}\left(\gamma_{\rm I}\sqrt{R_{\rm I}}+\sqrt{\gamma_{\rm I}\gamma_{\rm II} R_{\rm II}}\right), \label{eq:SudJum}
\eeq
where $\gamma_{\rm I,II}$  and $R_{\rm I,II}$ are respectively effective values for anisotropic gyromagnetic ratio\cite{supp,Maryasov:1982cg} and spin-flip rates for the relevant electron spin transitions of spins in site I and site II (see Sec.S6)\cite{supp}, $h$ is Planck's constant, $\mu_0$ is the vacuum permeability and $n$ is the density of interacting spins, which we assume to be $4.7\times 10^{17}\mathrm{cm}^{-3}$ for both sites. The two summed terms 
in \Eq{eq:SudJum} give separately the contributions to $T_\mathrm{SD}$ of a site I spin arising from spin flips in sites I and II. A more general description including other sources of line broadening is be discussed 
in Sec.S6\cite{supp}.

Values of $\gamma_{\rm I}\sim2.0\mu_B/h$ and $\gamma_{\rm II}\sim 6.0\mu_B/h$ can be extracted from the spin Hamiltonian, given the field magnitude and direction used for the data in \Fig{fig:eprddtem}, leading to dipole-coupling broadened linewidths on the order of $\sim200$~kHz (see details of analysis in Sec.S6\cite{supp}). 
Spin flip rates can be taken from the \tonee\ measurements shown above, using also the $g_{\rm eff}^2$-dependent one-phonon spin relaxation process which dominates at temperatures below 5~K (e.g.\ $R_{\rm I} = 180$~Hz and $R_{\rm II} = 3.7$~kHz at 4.5 K). The resulting prediction for $T_\mathrm{SD}$ using this model is shown in \Fig{fig:eprddtem} (solid red line) giving a good agreement with the Hahn echo data. To illustrate the effect of dynamics from the different sites, at 2.5~K we predict $T_\mathrm{SD,I}$ of 66~\microsec, made up of contributions of 650~\microsec\ and 73~\microsec\ from spin flips in site I and II, respectively. In this field orientation, spins in site II are more effective at driving spin decoherence by spectral diffusion due to their larger effective gyromagnetic ratio and shorter \tonee. In summary, spin-spin interactions between ions in different crystallographic sites play a key role in determining \ttwoe. 

\begin{figure}
	\centerline{\includegraphics[width=8.5cm]{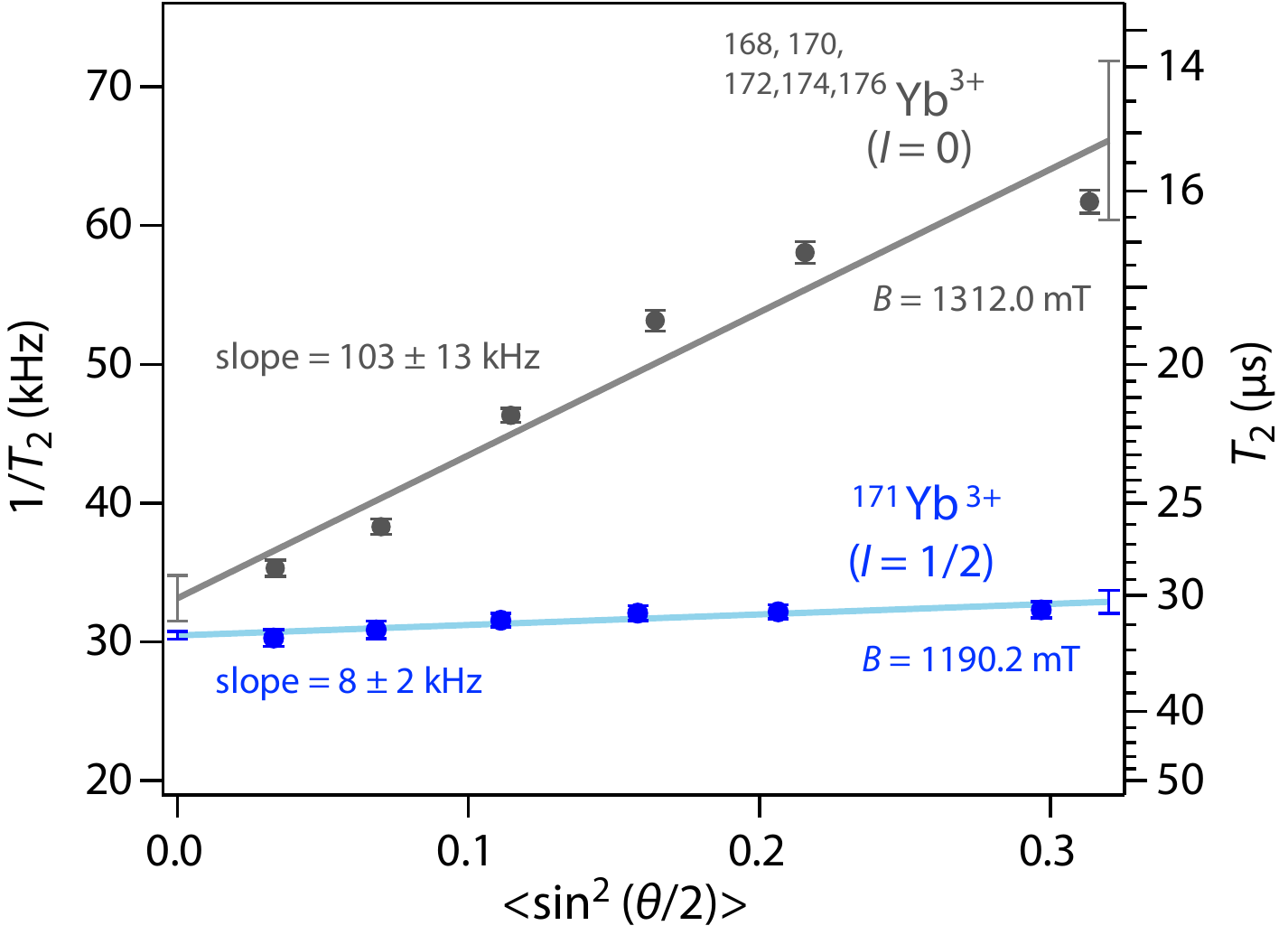}}
	\caption{The effects of instantaneous diffusion (ID) observed in measurements of \ttwoe\ can be seen by comparing results from the $I=0$ isotopes of Yb, and the ($I=1/2$) $^{171}\mathrm{Yb}$ isotope, which has a lower effective concentration of resonant spins. 
	$T_2$ was measured as a function of $\langle\sin^2 \theta/2\rangle$, where $\theta$ is rotation angle of the second pulse in the sequence ($\pi/2 - \tau - \theta - \tau - {\rm echo}$).
	Site II was used in these measurements, with the magnetic field $B$ applied along 65$^\circ$ from $D_1$ in the $D_1D_2$ plane.
	The duration of a $\pi$-pulse was 32~ns, and the temperature was 3~K.
	} \label{fig:ID}
\end{figure}

In the model above, the effect of instantaneous diffusion (ID) was ignored. ID can be viewed as a form of spectral diffusion induced by rotations of only those spins driven by the microwave pulses (`resonant spins')\cite{Boscaino:1992ef,Agnello:2001jr}. A key characteristic of ID is its dependence on the rotation angle $\theta$ of the second pulse in the spin-echo sequence:
\beq
T_{2,\rm ID}^{-1}=\frac{2 \pi^2 \mu_0 h n \gamma^2}{9 \sqrt 3} \left\langle\sin^2 \frac{\theta}{2}\right\rangle, \label{eq:instDiff}
\eeq
where $n$ here refers to concentration of \emph{resonant} spins which contribute to ID,\cite{Wolfowicz:2012foa} which can be far lower than the total spin concentration, especially in samples such as ours with multiple sites and hyperfine transitions.
$\langle\sin^2\theta/2\rangle$ is the average spin-flip probability achieved  by the second pulse in the echo sequence, bearing in mind the inhomogeneously broadened linewidth (FWHM $\sim 32$ MHz) and finite Rabi frequency ($\sim 15.6$ MHz in our set-up).\cite{Agnello:2001jr,Rakhmatullin:2009jfa,supp}
A $\theta$-dependence in the measured \ttwoe\ is therefore a signature of ID, which we 
identify using the ESR transition of $I=0$ isotopes, where the concentration of resonance spins is greatest (see \Fig{fig:ID}). In these experiments, $B$ is applied along 65$^\circ$ in $D_1D_2$ plane ($g_\mathrm{eff}\sim0.6$, $\gamma\sim2.1 \mu_B/h$\cite{supp}), and we compare the results for the $I=0$ isotopes, and the $^{171}\mathrm{Yb}$ isotope. The two measurements give the similar limit of around 30~\microsec\ for $\theta\rightarrow0$, determined by the spectral diffusion processes described above. 
However, while the \ttwoe\ measured for $^{171}\mathrm{Yb}$ shows a barely visible dependence on $\langle\sin^2 \frac{\theta}{2}\rangle$, the slope for the $I=0$ isotopes is a factor of $14 \pm 5$ larger, consistent with the increased concentration of resonant spins for that transition (a factor of 5 comes from the natural isotopic abundance, and a further factor of 2 from the hyperfine splitting in $^{171}\mathrm{Yb}$). Furthermore, assuming equal distribution of ions across the two sites, these slopes would suggest a total ion concentration of $38\pm5$ ppm, close to the expected value. 
 
\begin{figure}
   	\centerline{\includegraphics[width=8.5cm]{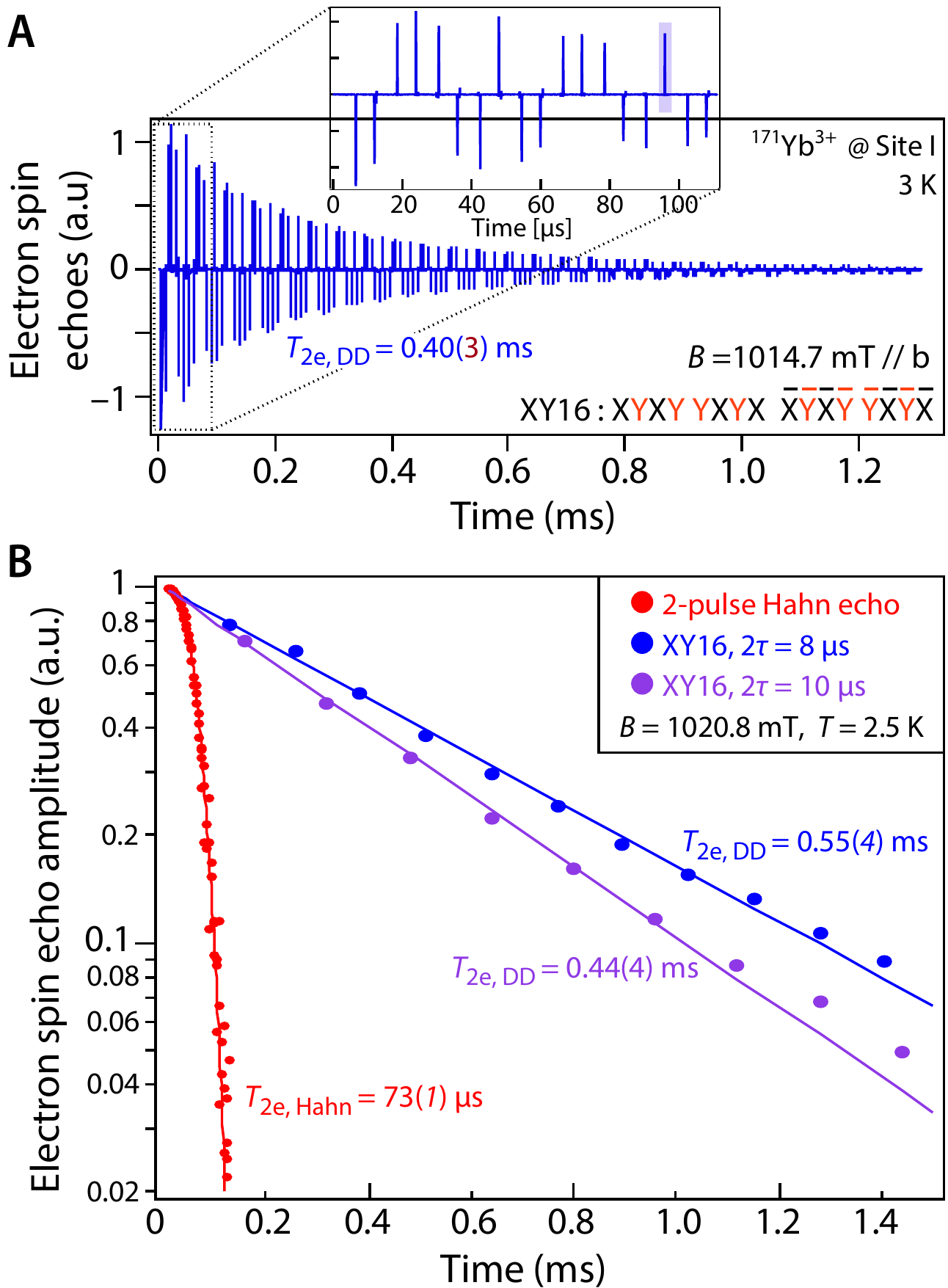}}
   	\caption{{\bf(a)} Time domain trace of spin echoes under XY16 dynamical decoupling, with the microwave control pulses suppressed by phase cycling.
	{\bf(b)} Echo amplitudes are shown as a function of total time from the initial $\pi/2$ pulse, comparing results from a 2-pulse (Hahn) echo and XY16. For the latter, the amplitude of the final echo in each 16-echo segment (highlighted in the inset of panel (a)) is shown, as are results for different values of $2\tau$ (the separation between each pulse). In contrast to the stretched exponential decay observed for 2-pulse \ttwoe\ measurements, it can be seen that $T_\mathrm{2, DD}$ follows an simple exponential decay. 	Experiments were performed using $^{171}\mathrm{Yb}$ in site I with the magnetic field $B$ as given applied along the $b$ crystal axis.} \label{fig:DD}
   \end{figure}

We have seen above that spectral diffusion from spin-flips of neighbouring spins forms a dominant contribution to the \ttwoe\ measured by a 2-pulse Hahn echo. Such effects can be mitigated through the application of dynamical decoupling (DD) schemes, such as the XY16 sequence\cite{Khodjasteh:2009ff,2012RSPTA.370.4748S}. The XY16 sequence is an example of a universal decoupling sequence (i.e.\ its performance is not a function of the initial spin state) and has good robustness to rotation angle errors in the pulses\cite{2012RSPTA.370.4748S}. \Fig{fig:DD} shows our results from applying concatenated XY16, as well as an illustration of the sequence itself, with the coherence time extended by dynamical decoupling to up to $T_\mathrm{DD}$ is 550 $\mu$s at 2.5~K. The effect of $2\tau$ (the separation in time between each pulse) is visible as a shorter $\tau$ is more effective at suppressing higher-frequency spectral diffusion\cite{Viola:book}, extending $T_\mathrm{2, DD}$, though $\tau\leq3$~\microsec\ could not be investigated for instrumental reasons. Nevertheless, dynamical decoupling is clearly an effective tool for the suppression of the effects of spectral diffusion on spin decoherence, with the resulting values for $T_\mathrm{2, DD}$ approaching the limit predicted by instantaneous diffusion (see \Fig{fig:eprddtem}), which XY16 is not able to effectively suppress.


\section{THREE-pulse echo measurements}\label{sec:SD}

    \begin{figure}
      	\centerline{\includegraphics[width=8.5cm]{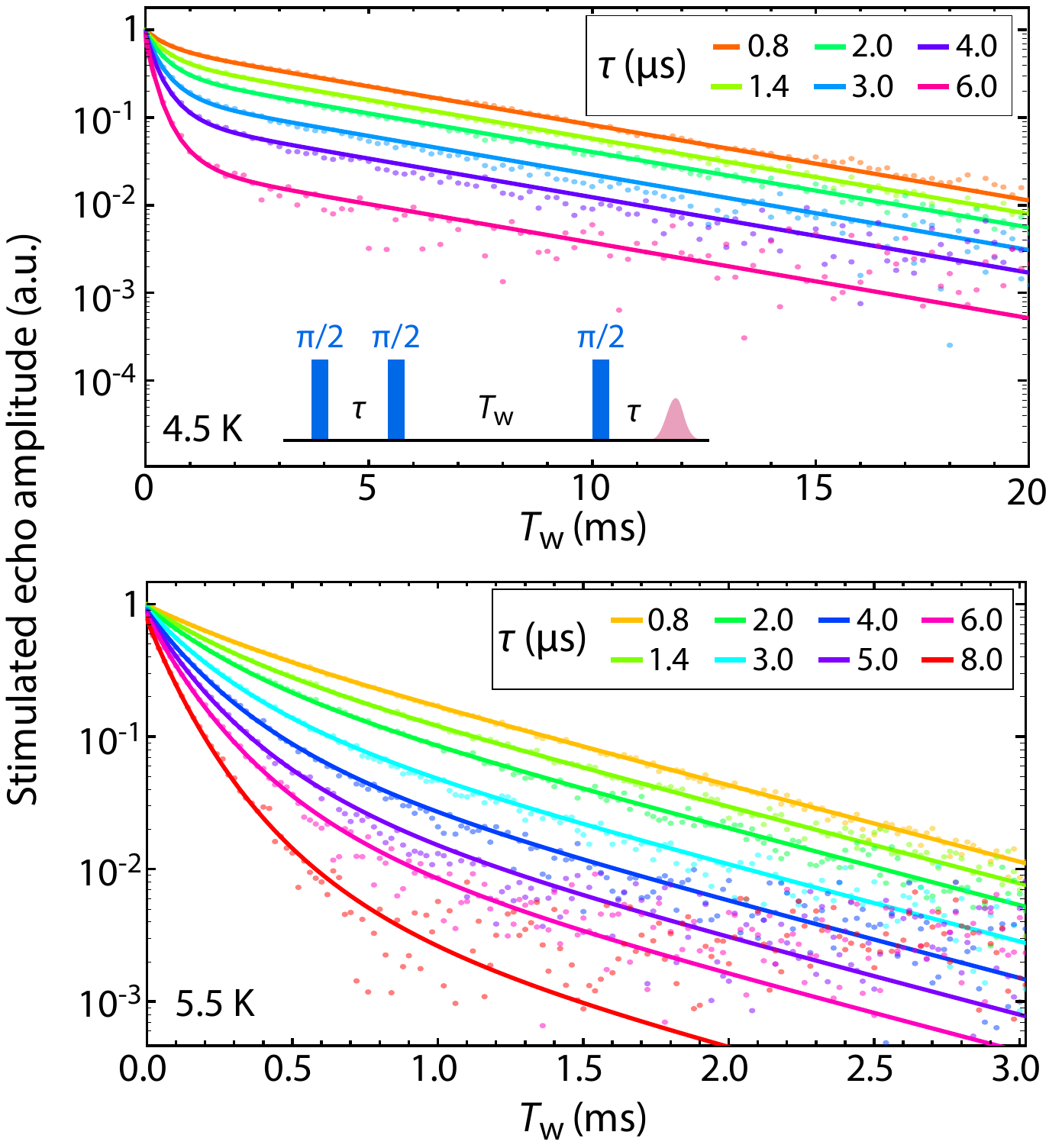}}
      	\caption{Three-pulse stimulated echo decays, as a function of $T_\mathrm{w}$ and for various $\tau$ (following the notation given in the inset) at 4.5 and 5.5~K.
	 These decays were fit (solid lines) using the model of Eqs.\ref{eq:STE1} and \ref{eq:STE2} with parameters shown in Table~\ref{tab:SD}, in order to study the effects of spectral diffusion. Measurements were taken using $^{171}\mathrm{Yb}^{3+}$ in site I at $B =1046.6$~mT aligned close to the $b$ axis.}
	\label{fig:ESD}
      \end{figure}
 
To verify our understanding of the role and strength of spectral diffusion in the \ttwoe\ measurements described above (for $T<6$~K), we use the three-pulse (stimulated) echo technique ($\frac{\pi}{2} - \tau - \frac{\pi}{2} - T_{\rm w} - \frac{\pi}{2} - \tau - {\rm echo}$)\cite{PhysRev.80.580,Hu:1978cj,schweiger2001principles}. The amplitude of the stimulated echo $A(\tau,T_{\rm w})$ decays according to a function of both the electron spin relaxation time \tonee, as well as the spectral diffusion linewidth $\Gamma_\mathrm{SD}$\cite{Bottger:2006jo,marino:pastel-00746050}:
\beq
  \frac{ A(\tau,T_w)}{A_0}=\exp\left[-\left(\frac{T_\mathrm w}{T_1}+2\pi\tau\Gamma_\mathrm{eff}\right)\right], 
  \label{eq:STE1}
\eeq
\beq
  {\rm where~}   \Gamma_\mathrm{eff}= \Gamma_0  + \frac{1}{2} \Gamma_\mathrm{SD}\left(R \tau + 1 - e^{-R T_\mathrm w}\right).
  \label{eq:STE2}
  \eeq
This formula is valid when $\tau$ is short compared with \tone, such that multiple  spin-flips are less probable.
Here, $\Gamma_0$ captures effects such as instantaneous diffusion and (single-ion) homogeneous broadening, while $R$ is the total spin flip rate. 
By measuring stimulated echo decay curves as a function of $T_{\rm w}$ for various values of $\tau$ (see \Fig{fig:ESD}) we obtain fitted values for $\Gamma_0$, $\Gamma_\mathrm{SD}$, $R$ and $T_1$, summarised in Table \ref{tab:SD}.


{ \renewcommand{\arraystretch}{1.5}
	\begin{table}
		\centering
		\begin{tabular}{|c|cccc|c|c|c|}
			\Xhline{3 \arrayrulewidth}
			\makecell{~Temp.~\\{[}K{]}} & \makecell{$\Gamma_0$\\ {~~[}kHz{]~~}} & \makecell{$\Gamma_\mathrm{SD}$\\ {~~[}kHz{]~~~}} & \makecell{R\\ {~~~[}kHz{]~~}}   & \makecell{$T_1^{-1}$\\ {~~[}kHz{]~~~}} & \makecell{$T^{-1}_\mathrm{1, IR}$\\{~~[}kHz{]~~}}   \\ \Xhline{3\arrayrulewidth}
			4.5	&  	3(1)		& 181(2) 	& 1.8(1)		& 0.203(2)		& 0.21 		\\ \hline
			5.5  & 	3.7(4)     & 192(4)      	& 2.3(1) 		& 1.33(1)     				& 1.24 	\\ \hline
			6.0  & 	4.4(3)    	& 187(6)      	& 3.2(1) 		& 3.66(3)     				& 3.58     \\
			\Xhline{3\arrayrulewidth}
		\end{tabular}
		\caption{Fitted values for parameters of spectral diffusion ($\Gamma_0, \Gamma_{\mathrm{SD}}, R, T_1^{-1}$) extrapolated from stimulated echo decay measurements (\Fig{fig:ESD}), as defined in the text. The fitted values of $T_1$ match well those of $T_{1, IR}$ obtained from inversion recovery experiments.} \label{tab:SD}
	\end{table}}
   
The extracted values for $T_1^{-1}$  match well those obtained from inversion recovery measurements, consistent with our assumption to neglect spin flip-flops.
The values for $\Gamma_\mathrm{SD}$ agree with the dipolar interactions of ions coupled by the effective value for anisotropic $\gamma$ (see Sec.S6\cite{supp}).
Values for $R$ are within a factor of two of those extracted from the analysis of 2-pulse echo decays described in the previous section (see Sec.S6\cite{supp}), and the remaining discrepancy could be due to our assumptions for estimating the \tone\ of spins in the other crystallographic site (for example, through effects such as cross-relaxation which have a weaker $B$ dependence than the single-phonon process). 
Finally, we note that $\Gamma_0$, though giving only a weak contribution in this temperature range leading to large errors bars, does appear larger at low temperatures than our expectations for instantaneous diffusion and homogenous broadening. We also note that  $T_{2, \rm DD}$ at 4.5~K is longer than the limit expected from $1/\pi \Gamma_0 \sim 0.1$~ms, suggesting that it captures an effect which can be suppressed by dynamical decoupling, such as an additional spectral diffusion mechanism with much slower dynamics. 

Another possible source of spectral diffusion not yet discussed arises from $^{89}\mathrm{Y}$ nuclear spins in the crystal. The $^{89}\mathrm{Y}$ nuclear spin flip-flop rate is known to be 8~Hz in the bulk \cite[Eq.15]{Bottger:2006jo}, and is expected to reduce to around 1.2 Hz for $^{89}\mathrm{Y}$ adjacent to the $\mathrm{Yb}^{3+}$ ion ($r = 3.39\angstrom$) due to the `frozen-core' effect\cite{marino:pastel-00746050
,Bottger:2006jo}. Such rates are lower than our extracted value of $R$ by two or three orders of magnitude, such that we can conclude that spectral diffusion from $^{89}\mathrm{Y}$ is not a major effect over the temperature range 4.5--6~K.


\section{Nuclear spin coherence}  \label{sec:NS}

We now move on to explore the coherent dynamics of the Yb nuclear spins. As stated above, the $^{171}\mathrm{Yb}$ nuclear spin is perhaps the most technologically interesting for optical interfaces due to its $I=1/2$ spin \cite{PhysRevA.60.2867}, however, as a preliminary study we focus here on the $^{173}\mathrm{Yb}$ ($I=5/2$) transitions because they occur at lower frequencies and are therefore technically easier to access using a typical ENDOR resonator. We expect our results on the coherence properties of $^{173}\mathrm{Yb}$ ENDOR transitions to provide a lower bound of expectations for $^{171}\mathrm{Yb}$, given that the former has only additional decoherence pathways due to its higher nuclear spin quantum number (see Fig.S2)\cite{supp}. We focus our studies on $B \parallel b$, where ENDOR frequencies of $^{173}\mathrm{Yb}$ in site I are less than 400 MHz and the degree of mixing 
 reduces the sensitivity of the nuclear spin transitions to magnetic field fluctuations. An example Davies ENDOR spectrum is shown in \Fig{fig:spectra}(d), and our coherence time measurements are performed on the $m_I=-1/2^-:+3/2^-$ transition (see \Fig{fig:spectra}(a)) which had a frequency of  223~MHz $B= 989$~mT.

\begin{figure}
    	\centerline{\includegraphics[width=8.5cm]{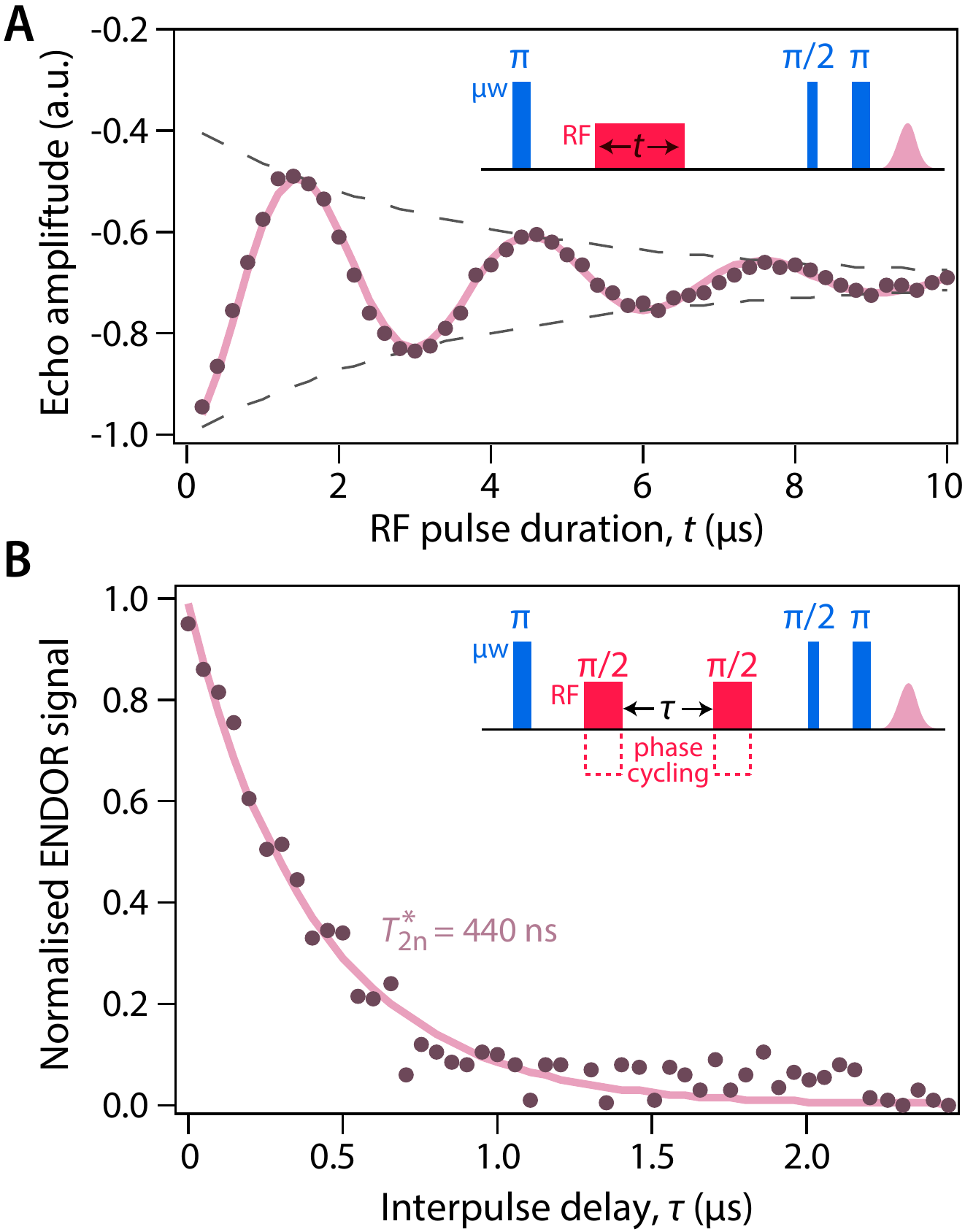}}
    	\caption{\textbf{(a)} Rabi oscillations of the $m_I=-1/2^-:+3/2^-$ transition of $^{173}\mathrm{Yb}^{3+}$ in site I at 4.5~K, $B= 989$~mT~$\parallel b$, with an RF frequency of 223~MHz (see \Fig{fig:spectra}(a)). 
	\textbf{(b)} $T_{2,N}^*$ measurements to study the inhomogeneous linewidth ($1/\pi T_{2}^*$) of nuclear spins.
	The horizontal axis is a time period between two 90$^\circ$ RF pulses. 4-step phase cycling on the RF pulses was used to give zero offset.}
	 \label{fig:NS}
    \end{figure}

Rabi oscillation measurements were first performed (Fig.\ref{fig:NS}a) to find the optimum pulse durations for the nuclear spins, using a sequence based on Davies ENDOR as shown in the inset. Based on these measurements, we chose an RF duration of 1.5~\microsec\ for a $\pi$-pulse, using a 100~W amplifier with 40\% gain.
A ``Tidy" RF $\pi$-pulse\cite{MORTON2008315} was applied at the end of each sequence to mitigate the effects of slow nuclear spin relaxation. We studied ENDOR signals without and without the Tidy pulse at various shot repetitions times, in order to estimate nuclear spin and cross-relaxation times, which we find to be around 4 times longer than the pure electron spin relaxation times \tonee\ (see Sec.S7\cite{supp}). This is consistent with the large dipole moment of this ENDOR transition (estimated to be 0.04 $\mu_B$) resulting from state mixing.  

We measured $T_{2n}^*$ using a RF Ramsey pulse sequence ($\frac{\pi}{2}-\tau-\frac{\pi}{2}$) replacing the usual RF $\pi$-pulse in a Davies ENDOR measurement (see \Fig{fig:NS}b), making use of 4-step phase cycling\cite{Fauth:1986jt} in the RF pulses to produce a zero baseline. The resulting exponential decay with $T_{2n}^*=440$~ns is consistent with the Lorentzian lineshape of width 0.72~MHz see in \Fig{fig:spectra}(c). Next, we measure \ttwon\ using a nuclear spin echo measurement~, adding an additional RF $\pi$-pulse to the previous experiment\cite{Pla:2013io}. As for the electron spin coherence, we find a stretched exponential decay with a stretch factor of 1.7. However, \ttwon\ is 0.35(2)~ms at 4.5~K, which is 14 times longer than \ttwoe~(25~\microsec) and twice as long as $T_\mathrm{2e, DD}$, at the same temperature.

\begin{figure}[t]
    	\includegraphics[width=8.5cm]{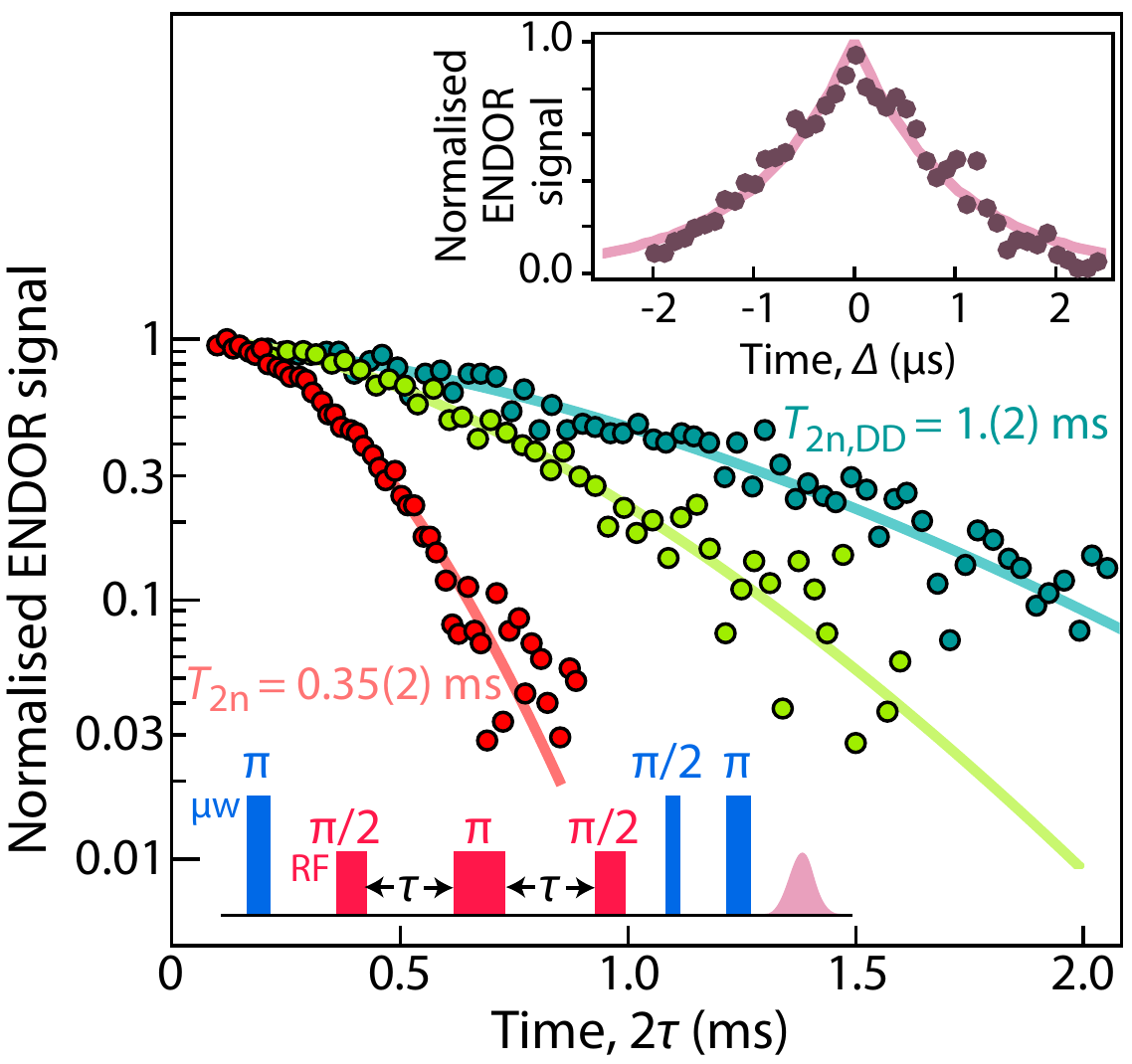}
    	\caption{Nuclear spin echo decay as a function of $\tau$, using sequence shown\cite{Pla:2013io}, fit to a stretched exponential decay with $T_{2n} = 0.35$ ms and stretch factor 1.7 (red solid curve). Measurements were performed using an RF frequency of 223~MHz, addressing the $m_I=-1/2^-:+3/2^-$ transition of $^{173}\mathrm{Yb}^{3+}$ in site I, $B= 989$~mT~$\parallel b$. 4-step phase cycling was used to give zero offet. Inset shows a typical nuclear spin echo measured by sweeping the delay ($\tau+\Delta$) between the final two RF pulses, with $\tau=5$~\microsec. The dynamic decoupling scheme of XY16 gave futher improvements of slow decays. The characteristic time ($T_{2n,DD}$) increased to $0.72$ ms in a cycle of XY16 (green), and $1.2$ ms with a stretch factor of 1.6 in two cycles of XY16 (blue).}
	 \label{fig:nucHahn}
    \end{figure}

To understand the nuclear spin coherence times we can use the same spectral diffusion model used for the electron spin (\Sec{sec:EPRDD}, Eq.\ref{eq:SudJum}). Still assuming $n=4.7\times 10^{17}$ cm$^{-3}$, and using the smaller gyromagnetic ratio of nuclear spin (for this transition, $h\gamma_{N}/\mu_B \sim 0.01(5)$), we expect the nuclear spin $T_\mathrm{SD}$ to be $0.6\pm0.2$~ms, in good agreement with the measured value. As for the electron spin, the measured nuclear spin coherence could be increased using DD, increasing to $0.7 \pm 0.3$ ms and $1.2 \pm 0.2$ ms under one and two cycles of  XY16, respectively. A stretched exponential decay remained visible under the DD, providing evidence of limited ability of such nuclear DD to suppress spectral diffusion in for this doping concentration.

\section*{Conclusions}                      

We have studied the $\mathrm{Yb}^{3+}$:YSO electron and nuclear spin relaxation and decoherence dynamics, along with the mechanisms behind them. The spin relaxation times of the electron spin are governed by a one-phonon process at temperature below 4~K, and we have measured times up to 5~s at 2.5~K for site II with $B\parallel b$, where the high g-factor enables X-band ESR at low magnetic fields. We would expect \tonee\ to continue to rise as the temperature is reduced, reaching a limit of about 50~s at temperatures $\leq100$~mK for the same ESR transition, relevant for efforts to couple Yb:YSO spins to superconducting resonators in dilution refrigerators\cite{Bushev:2011be}. We find the nuclear spin relaxation times \tonen\ are a small factor longer than \tonee\, consistent with the significant degree of spin mixing.

Coherence lifetimes for the electron and nuclear spin are dominated by spectral diffusion, predominantly from other $\mathrm{Yb}^{3+}$ spins occupying either site. Effects from $^{89}$Y nuclear spins appear negligible in the temperature regime studied here, and we do not see evidence of a significant contribution from other paramagnetic impurities which might have been found in the YSO host material. At 4.5~K, we measure \ttwoe~$=25$~\microsec\ and \ttwon~$=350$~\microsec\ for $^{173}\mathrm{Yb}$ in site I and $g_\mathrm{eff} = 0.7$, which are similar in magnitude to those measured for Nd:YSO\cite{Wolfowicz:2015exa,Sukhanov:2017gx}. Using XY16 dynamical decoupling, we can largely suppress the effects of spectral diffusion, leading to coherence lifetimes of $T_\mathrm{2e, DD} = 0.55 $ ms at 2.5~K, which may be limited by instantaneous diffusion.

For future studies on $^{171}\mathrm{Yb}$, there are significant advantages to using isotopically enriched Yb to dope the YSO. A doping level of around 7~ppm would give an equivalent ESR signal to that obtained in our sample of (50~ppm) natural abundance Yb:YSO, but with a substantial reduction in spectral diffusion. Lowering the doping level further would extend the limit of instantaneous diffusion and electron spin coherence times in the milliseconds should be achievable.

%



\begin{thebibliography}{63}%
	\makeatletter
	\providecommand \@ifxundefined [1]{%
		\@ifx{#1\undefined}
	}%
	\providecommand \@ifnum [1]{%
		\ifnum #1\expandafter \@firstoftwo
		\else \expandafter \@secondoftwo
		\fi
	}%
	\providecommand \@ifx [1]{%
		\ifx #1\expandafter \@firstoftwo
		\else \expandafter \@secondoftwo
		\fi
	}%
	\providecommand \natexlab [1]{#1}%
	\providecommand \enquote  [1]{``#1''}%
	\providecommand \bibnamefont  [1]{#1}%
	\providecommand \bibfnamefont [1]{#1}%
	\providecommand \citenamefont [1]{#1}%
	\providecommand \href@noop [0]{\@secondoftwo}%
	\providecommand \href [0]{\begingroup \@sanitize@url \@href}%
	\providecommand \@href[1]{\@@startlink{#1}\@@href}%
	\providecommand \@@href[1]{\endgroup#1\@@endlink}%
	\providecommand \@sanitize@url [0]{\catcode `\\12\catcode `\$12\catcode
		`\&12\catcode `\#12\catcode `\^12\catcode `\_12\catcode `\%12\relax}%
	\providecommand \@@startlink[1]{}%
	\providecommand \@@endlink[0]{}%
	\providecommand \url  [0]{\begingroup\@sanitize@url \@url }%
	\providecommand \@url [1]{\endgroup\@href {#1}{\urlprefix }}%
	\providecommand \urlprefix  [0]{URL }%
	\providecommand \Eprint [0]{\href }%
	\providecommand \doibase [0]{http://dx.doi.org/}%
	\providecommand \selectlanguage [0]{\@gobble}%
	\providecommand \bibinfo  [0]{\@secondoftwo}%
	\providecommand \bibfield  [0]{\@secondoftwo}%
	\providecommand \translation [1]{[#1]}%
	\providecommand \BibitemOpen [0]{}%
	\providecommand \bibitemStop [0]{}%
	\providecommand \bibitemNoStop [0]{.\EOS\space}%
	\providecommand \EOS [0]{\spacefactor3000\relax}%
	\providecommand \BibitemShut  [1]{\csname bibitem#1\endcsname}%
	\let\auto@bib@innerbib\@empty
	\bibitem [{\citenamefont {Goldner}\ \emph {et~al.}(2015)\citenamefont
		{Goldner}, \citenamefont {Ferrier},\ and\ \citenamefont
		{Guillot-No\"el}}]{Goldner:2015}%
	\BibitemOpen
	\bibfield  {author} {\bibinfo {author} {\bibfnamefont {P.}~\bibnamefont
			{Goldner}}, \bibinfo {author} {\bibfnamefont {A.}~\bibnamefont {Ferrier}}, \
		and\ \bibinfo {author} {\bibfnamefont {O.}~\bibnamefont {Guillot-No\"el}},\
	}in\ \href@noop {} {\emph {\bibinfo {booktitle} {Handbook on the Physics and
				Chemistry of Rare Earths}}},\ Vol.~\bibinfo {volume} {46},\ \bibinfo {editor}
	{edited by\ \bibinfo {editor} {\bibfnamefont {J.-C.~G.}\ \bibnamefont
			{B\"unzli}}\ and\ \bibinfo {editor} {\bibfnamefont {V.~K.}\ \bibnamefont
			{Pecharsky}}}\ (\bibinfo  {publisher} {Eds. Amsterdam: Elsevier},\ \bibinfo
	{year} {2015})\ pp.\ \bibinfo {pages} {1--78}\BibitemShut {NoStop}%
	\bibitem [{\citenamefont {Afzelius}\ \emph {et~al.}(2013)\citenamefont
		{Afzelius}, \citenamefont {Sangouard}, \citenamefont {Johansson},
		\citenamefont {Staudt},\ and\ \citenamefont {Wilson}}]{Afzelius:2013}%
	\BibitemOpen
	\bibfield  {author} {\bibinfo {author} {\bibfnamefont {M.}~\bibnamefont
			{Afzelius}}, \bibinfo {author} {\bibfnamefont {N.}~\bibnamefont {Sangouard}},
		\bibinfo {author} {\bibfnamefont {G.}~\bibnamefont {Johansson}}, \bibinfo
		{author} {\bibfnamefont {M.~U.}\ \bibnamefont {Staudt}}, \ and\ \bibinfo
		{author} {\bibfnamefont {C.~M.}\ \bibnamefont {Wilson}},\ }\href
	{http://stacks.iop.org/1367-2630/15/i=6/a=065008} {\bibfield  {journal}
		{\bibinfo  {journal} {New J. Phys.}\ }\textbf {\bibinfo {volume} {15}},\
		\bibinfo {pages} {065008} (\bibinfo {year} {2013})}\BibitemShut {NoStop}%
	\bibitem [{\citenamefont {Probst}\ \emph {et~al.}(2013)\citenamefont {Probst},
		\citenamefont {Rotzinger}, \citenamefont {W{\"u}nsch}, \citenamefont {Jung},
		\citenamefont {Jerger}, \citenamefont {Siegel}, \citenamefont {Ustinov},\
		and\ \citenamefont {Bushev}}]{Probst:2013vp}%
	\BibitemOpen
	\bibfield  {author} {\bibinfo {author} {\bibfnamefont {S.}~\bibnamefont
			{Probst}}, \bibinfo {author} {\bibfnamefont {H.}~\bibnamefont {Rotzinger}},
		\bibinfo {author} {\bibfnamefont {S.}~\bibnamefont {W{\"u}nsch}}, \bibinfo
		{author} {\bibfnamefont {P.}~\bibnamefont {Jung}}, \bibinfo {author}
		{\bibfnamefont {M.}~\bibnamefont {Jerger}}, \bibinfo {author} {\bibfnamefont
			{M.}~\bibnamefont {Siegel}}, \bibinfo {author} {\bibfnamefont {A.~V.}\
			\bibnamefont {Ustinov}}, \ and\ \bibinfo {author} {\bibfnamefont {P.~A.}\
			\bibnamefont {Bushev}},\ }\href
	{http://link.aps.org/doi/10.1103/PhysRevLett.110.157001} {\bibfield
		{journal} {\bibinfo  {journal} {Phys. Rev. Lett.}\ }\textbf {\bibinfo
			{volume} {110}},\ \bibinfo {pages} {157001} (\bibinfo {year}
		{2013})}\BibitemShut {NoStop}%
	\bibitem [{\citenamefont {Williamson}\ \emph {et~al.}(2014)\citenamefont
		{Williamson}, \citenamefont {Chen},\ and\ \citenamefont
		{Longdell}}]{Williamson:2014fb}%
	\BibitemOpen
	\bibfield  {author} {\bibinfo {author} {\bibfnamefont {L.~A.}\ \bibnamefont
			{Williamson}}, \bibinfo {author} {\bibfnamefont {Y.-H.}\ \bibnamefont
			{Chen}}, \ and\ \bibinfo {author} {\bibfnamefont {J.~J.}\ \bibnamefont
			{Longdell}},\ }\href {\doibase 10.1103/PhysRevLett.113.203601} {\bibfield
		{journal} {\bibinfo  {journal} {Phys. Rev. Lett.}\ }\textbf {\bibinfo
			{volume} {113}},\ \bibinfo {pages} {203601} (\bibinfo {year}
		{2014})}\BibitemShut {NoStop}%
	\bibitem [{\citenamefont {Wolfowicz}\ \emph {et~al.}(2015)\citenamefont
		{Wolfowicz}, \citenamefont {Maier-Flaig}, \citenamefont {Marino},
		\citenamefont {Ferrier}, \citenamefont {Vezin}, \citenamefont {Morton},\ and\
		\citenamefont {Goldner}}]{Wolfowicz:2015exa}%
	\BibitemOpen
	\bibfield  {author} {\bibinfo {author} {\bibfnamefont {G.}~\bibnamefont
			{Wolfowicz}}, \bibinfo {author} {\bibfnamefont {H.}~\bibnamefont
			{Maier-Flaig}}, \bibinfo {author} {\bibfnamefont {R.}~\bibnamefont {Marino}},
		\bibinfo {author} {\bibfnamefont {A.}~\bibnamefont {Ferrier}}, \bibinfo
		{author} {\bibfnamefont {H.}~\bibnamefont {Vezin}}, \bibinfo {author}
		{\bibfnamefont {J.~J.~L.}\ \bibnamefont {Morton}}, \ and\ \bibinfo {author}
		{\bibfnamefont {P.}~\bibnamefont {Goldner}},\ }\href {\doibase
		10.1103/PhysRevLett.114.170503} {\bibfield  {journal} {\bibinfo  {journal}
			{Phys. Rev. Lett.}\ }\textbf {\bibinfo {volume} {114}},\ \bibinfo {pages}
		{170503} (\bibinfo {year} {2015})}\BibitemShut {NoStop}%
	\bibitem [{\citenamefont {Rancic}\ \emph {et~al.}(2017)\citenamefont {Rancic},
		\citenamefont {Hedges}, \citenamefont {Ahlefeldt},\ and\ \citenamefont
		{Sellars}}]{Rancic:2017bn}%
	\BibitemOpen
	\bibfield  {author} {\bibinfo {author} {\bibfnamefont {M.}~\bibnamefont
			{Rancic}}, \bibinfo {author} {\bibfnamefont {M.~P.}\ \bibnamefont {Hedges}},
		\bibinfo {author} {\bibfnamefont {R.~L.}\ \bibnamefont {Ahlefeldt}}, \ and\
		\bibinfo {author} {\bibfnamefont {M.~J.}\ \bibnamefont {Sellars}},\ }\href
	{https://www.nature.com/articles/nphys4254} {\bibfield  {journal} {\bibinfo
			{journal} {Nat. Phys.}\ }\textbf {\bibinfo {volume} {5}},\ \bibinfo {pages}
		{nphys4254} (\bibinfo {year} {2017})}\BibitemShut {NoStop}%
	\bibitem [{\citenamefont {Tittel}\ \emph {et~al.}(2010)\citenamefont {Tittel},
		\citenamefont {Afzelius}, \citenamefont {Chaneli{\`e}re}, \citenamefont
		{Cone}, \citenamefont {Kr{\"o}ll}, \citenamefont {Moiseev},\ and\
		\citenamefont {Sellars}}]{Tittel:2010bp}%
	\BibitemOpen
	\bibfield  {author} {\bibinfo {author} {\bibfnamefont {W.}~\bibnamefont
			{Tittel}}, \bibinfo {author} {\bibfnamefont {M.}~\bibnamefont {Afzelius}},
		\bibinfo {author} {\bibfnamefont {T.}~\bibnamefont {Chaneli{\`e}re}},
		\bibinfo {author} {\bibfnamefont {R.~L.}\ \bibnamefont {Cone}}, \bibinfo
		{author} {\bibfnamefont {S.}~\bibnamefont {Kr{\"o}ll}}, \bibinfo {author}
		{\bibfnamefont {S.~A.}\ \bibnamefont {Moiseev}}, \ and\ \bibinfo {author}
		{\bibfnamefont {M.~J.}\ \bibnamefont {Sellars}},\ }\href
	{http://onlinelibrary.wiley.com/doi/10.1002/lpor.200810056/ abstract}
	{\bibfield  {journal} {\bibinfo  {journal} {Laser {\&} Photon. Rev.}\
		}\textbf {\bibinfo {volume} {4}},\ \bibinfo {pages} {244} (\bibinfo {year}
		{2010})}\BibitemShut {NoStop}%
	\bibitem [{\citenamefont {Bussi{\`e}res}\ \emph {et~al.}(2014)\citenamefont
		{Bussi{\`e}res}, \citenamefont {Clausen}, \citenamefont {Tiranov},
		\citenamefont {Korzh}, \citenamefont {Verma}, \citenamefont {Nam},
		\citenamefont {Marsili}, \citenamefont {Ferrier}, \citenamefont {Goldner},
		\citenamefont {Herrmann}, \citenamefont {Silberhorn}, \citenamefont {Sohler},
		\citenamefont {Afzelius},\ and\ \citenamefont {Gisin}}]{Bussieres:2014dc}%
	\BibitemOpen
	\bibfield  {author} {\bibinfo {author} {\bibfnamefont {F.}~\bibnamefont
			{Bussi{\`e}res}}, \bibinfo {author} {\bibfnamefont {C.}~\bibnamefont
			{Clausen}}, \bibinfo {author} {\bibfnamefont {A.}~\bibnamefont {Tiranov}},
		\bibinfo {author} {\bibfnamefont {B.}~\bibnamefont {Korzh}}, \bibinfo
		{author} {\bibfnamefont {V.~B.}\ \bibnamefont {Verma}}, \bibinfo {author}
		{\bibfnamefont {S.~W.}\ \bibnamefont {Nam}}, \bibinfo {author} {\bibfnamefont
			{F.}~\bibnamefont {Marsili}}, \bibinfo {author} {\bibfnamefont
			{A.}~\bibnamefont {Ferrier}}, \bibinfo {author} {\bibfnamefont
			{P.}~\bibnamefont {Goldner}}, \bibinfo {author} {\bibfnamefont
			{H.}~\bibnamefont {Herrmann}}, \bibinfo {author} {\bibfnamefont
			{C.}~\bibnamefont {Silberhorn}}, \bibinfo {author} {\bibfnamefont
			{W.}~\bibnamefont {Sohler}}, \bibinfo {author} {\bibfnamefont
			{M.}~\bibnamefont {Afzelius}}, \ and\ \bibinfo {author} {\bibfnamefont
			{N.}~\bibnamefont {Gisin}},\ }\href {\doibase 10.1038/nphoton.2014.215}
	{\bibfield  {journal} {\bibinfo  {journal} {Nat. Photon.}\ }\textbf {\bibinfo
			{volume} {8}},\ \bibinfo {pages} {775} (\bibinfo {year} {2014})}\BibitemShut
	{NoStop}%
	\bibitem [{\citenamefont {Saglamyurek}\ \emph {et~al.}(2015)\citenamefont
		{Saglamyurek}, \citenamefont {Jin}, \citenamefont {Verma}, \citenamefont
		{Shaw}, \citenamefont {Marsili}, \citenamefont {Nam}, \citenamefont {Oblak},\
		and\ \citenamefont {Tittel}}]{Saglamyurek:2015esa}%
	\BibitemOpen
	\bibfield  {author} {\bibinfo {author} {\bibfnamefont {E.}~\bibnamefont
			{Saglamyurek}}, \bibinfo {author} {\bibfnamefont {J.}~\bibnamefont {Jin}},
		\bibinfo {author} {\bibfnamefont {V.~B.}\ \bibnamefont {Verma}}, \bibinfo
		{author} {\bibfnamefont {M.~D.}\ \bibnamefont {Shaw}}, \bibinfo {author}
		{\bibfnamefont {F.}~\bibnamefont {Marsili}}, \bibinfo {author} {\bibfnamefont
			{S.~W.}\ \bibnamefont {Nam}}, \bibinfo {author} {\bibfnamefont
			{D.}~\bibnamefont {Oblak}}, \ and\ \bibinfo {author} {\bibfnamefont
			{W.}~\bibnamefont {Tittel}},\ }\href
	{https://www.nature.com/articles/nphoton.2014.311} {\bibfield  {journal}
		{\bibinfo  {journal} {Nat. Photonics}\ }\textbf {\bibinfo {volume} {9}},\
		\bibinfo {pages} {83} (\bibinfo {year} {2015})}\BibitemShut {NoStop}%
	\bibitem [{\citenamefont {Zhong}\ \emph {et~al.}(2015)\citenamefont {Zhong},
		\citenamefont {Hedges}, \citenamefont {Ahlefeldt}, \citenamefont
		{Bartholomew}, \citenamefont {Beavan}, \citenamefont {Wittig}, \citenamefont
		{Longdell},\ and\ \citenamefont {Sellars}}]{Zhong:2015bw}%
	\BibitemOpen
	\bibfield  {author} {\bibinfo {author} {\bibfnamefont {M.}~\bibnamefont
			{Zhong}}, \bibinfo {author} {\bibfnamefont {M.~P.}\ \bibnamefont {Hedges}},
		\bibinfo {author} {\bibfnamefont {R.~L.}\ \bibnamefont {Ahlefeldt}}, \bibinfo
		{author} {\bibfnamefont {J.~G.}\ \bibnamefont {Bartholomew}}, \bibinfo
		{author} {\bibfnamefont {S.~E.}\ \bibnamefont {Beavan}}, \bibinfo {author}
		{\bibfnamefont {S.~M.}\ \bibnamefont {Wittig}}, \bibinfo {author}
		{\bibfnamefont {J.~J.}\ \bibnamefont {Longdell}}, \ and\ \bibinfo {author}
		{\bibfnamefont {M.~J.}\ \bibnamefont {Sellars}},\ }\href {\doibase
		10.1038/nature14025} {\bibfield  {journal} {\bibinfo  {journal} {Nature}\
		}\textbf {\bibinfo {volume} {517}},\ \bibinfo {pages} {177} (\bibinfo {year}
		{2015})}\BibitemShut {NoStop}%
	\bibitem [{\citenamefont {Jobez}\ \emph {et~al.}(2015)\citenamefont {Jobez},
		\citenamefont {Laplane}, \citenamefont {Timoney}, \citenamefont {Gisin},
		\citenamefont {Ferrier}, \citenamefont {Goldner},\ and\ \citenamefont
		{Afzelius}}]{Jobez:2015gt}%
	\BibitemOpen
	\bibfield  {author} {\bibinfo {author} {\bibfnamefont {P.}~\bibnamefont
			{Jobez}}, \bibinfo {author} {\bibfnamefont {C.}~\bibnamefont {Laplane}},
		\bibinfo {author} {\bibfnamefont {N.}~\bibnamefont {Timoney}}, \bibinfo
		{author} {\bibfnamefont {N.}~\bibnamefont {Gisin}}, \bibinfo {author}
		{\bibfnamefont {A.}~\bibnamefont {Ferrier}}, \bibinfo {author} {\bibfnamefont
			{P.}~\bibnamefont {Goldner}}, \ and\ \bibinfo {author} {\bibfnamefont
			{M.}~\bibnamefont {Afzelius}},\ }\href
	{https://journals.aps.org/prl/abstract/10.1103/ PhysRevLett.114.230502}
	{\bibfield  {journal} {\bibinfo  {journal} {Phys. Rev. Lett.}\ }\textbf
		{\bibinfo {volume} {114}},\ \bibinfo {pages} {230502} (\bibinfo {year}
		{2015})}\BibitemShut {NoStop}%
	\bibitem [{\citenamefont {Kutluer}\ \emph {et~al.}(2017)\citenamefont
		{Kutluer}, \citenamefont {Mazzera},\ and\ \citenamefont
		{de~Riedmatten}}]{Kutluer:2017ci}%
	\BibitemOpen
	\bibfield  {author} {\bibinfo {author} {\bibfnamefont {K.}~\bibnamefont
			{Kutluer}}, \bibinfo {author} {\bibfnamefont {M.}~\bibnamefont {Mazzera}}, \
		and\ \bibinfo {author} {\bibfnamefont {H.}~\bibnamefont {de~Riedmatten}},\
	}\href {https://journals.aps.org/prl/abstract/10.1103/
		PhysRevLett.118.210502} {\bibfield  {journal} {\bibinfo  {journal} {Phys.
				Rev. Lett.}\ }\textbf {\bibinfo {volume} {118}},\ \bibinfo {pages} {210502}
		(\bibinfo {year} {2017})}\BibitemShut {NoStop}%
	\bibitem [{\citenamefont {Chen}\ \emph {et~al.}(2016)\citenamefont {Chen},
		\citenamefont {Fernandez-Gonzalvo},\ and\ \citenamefont
		{Longdell}}]{Chen:2016uy}%
	\BibitemOpen
	\bibfield  {author} {\bibinfo {author} {\bibfnamefont {Y.-H.}\ \bibnamefont
			{Chen}}, \bibinfo {author} {\bibfnamefont {X.}~\bibnamefont
			{Fernandez-Gonzalvo}}, \ and\ \bibinfo {author} {\bibfnamefont {J.~J.}\
			\bibnamefont {Longdell}},\ }\href
	{http://link.aps.org/doi/10.1103/PhysRevB.94.075117} {\bibfield  {journal}
		{\bibinfo  {journal} {Phys. Rev. B}\ }\textbf {\bibinfo {volume} {94}},\
		\bibinfo {pages} {075117} (\bibinfo {year} {2016})}\BibitemShut {NoStop}%
	\bibitem [{\citenamefont {O{\textquoteright}Brien}\ \emph
		{et~al.}(2014)\citenamefont {O{\textquoteright}Brien}, \citenamefont {Lauk},
		\citenamefont {Blum}, \citenamefont {Morigi},\ and\ \citenamefont
		{Fleischhauer}}]{OBrien:2014dc}%
	\BibitemOpen
	\bibfield  {author} {\bibinfo {author} {\bibfnamefont {C.}~\bibnamefont
			{O{\textquoteright}Brien}}, \bibinfo {author} {\bibfnamefont
			{N.}~\bibnamefont {Lauk}}, \bibinfo {author} {\bibfnamefont {S.}~\bibnamefont
			{Blum}}, \bibinfo {author} {\bibfnamefont {G.}~\bibnamefont {Morigi}}, \ and\
		\bibinfo {author} {\bibfnamefont {M.}~\bibnamefont {Fleischhauer}},\ }\href
	{\doibase 10.1103/PhysRevLett.113.063603} {\bibfield  {journal} {\bibinfo
			{journal} {Phys. Rev. Lett.}\ }\textbf {\bibinfo {volume} {113}},\ \bibinfo
		{pages} {063603} (\bibinfo {year} {2014})}\BibitemShut {NoStop}%
	\bibitem [{\citenamefont {Dajczgewand}\ \emph {et~al.}(2014)\citenamefont
		{Dajczgewand}, \citenamefont {Le~Gou{\"e}t}, \citenamefont
		{Louchet-Chauvet},\ and\ \citenamefont
		{Chaneli{\`e}re}}]{Dajczgewand:2014et}%
	\BibitemOpen
	\bibfield  {author} {\bibinfo {author} {\bibfnamefont {J.}~\bibnamefont
			{Dajczgewand}}, \bibinfo {author} {\bibfnamefont {J.-L.}\ \bibnamefont
			{Le~Gou{\"e}t}}, \bibinfo {author} {\bibfnamefont {A.}~\bibnamefont
			{Louchet-Chauvet}}, \ and\ \bibinfo {author} {\bibfnamefont {T.}~\bibnamefont
			{Chaneli{\`e}re}},\ }\href
	{https://www.osapublishing.org/ol/abstract.cfm?uri=ol-39-9- 2711} {\bibfield
		{journal} {\bibinfo  {journal} {Opt. Lett.}\ }\textbf {\bibinfo {volume}
			{39}},\ \bibinfo {pages} {2711} (\bibinfo {year} {2014})}\BibitemShut
	{NoStop}%
	\bibitem [{\citenamefont {Zhong}\ \emph {et~al.}(2017)\citenamefont {Zhong},
		\citenamefont {Kindem}, \citenamefont {Bartholomew}, \citenamefont {Rochman},
		\citenamefont {Craiciu}, \citenamefont {Miyazono}, \citenamefont
		{Bettinelli}, \citenamefont {Cavalli}, \citenamefont {Verma}, \citenamefont
		{Nam}, \citenamefont {Marsili}, \citenamefont {Shaw}, \citenamefont {Beyer},\
		and\ \citenamefont {Faraon}}]{Zhong:2017fe}%
	\BibitemOpen
	\bibfield  {author} {\bibinfo {author} {\bibfnamefont {T.}~\bibnamefont
			{Zhong}}, \bibinfo {author} {\bibfnamefont {J.~M.}\ \bibnamefont {Kindem}},
		\bibinfo {author} {\bibfnamefont {J.~G.}\ \bibnamefont {Bartholomew}},
		\bibinfo {author} {\bibfnamefont {J.}~\bibnamefont {Rochman}}, \bibinfo
		{author} {\bibfnamefont {I.}~\bibnamefont {Craiciu}}, \bibinfo {author}
		{\bibfnamefont {E.}~\bibnamefont {Miyazono}}, \bibinfo {author}
		{\bibfnamefont {M.}~\bibnamefont {Bettinelli}}, \bibinfo {author}
		{\bibfnamefont {E.}~\bibnamefont {Cavalli}}, \bibinfo {author} {\bibfnamefont
			{V.}~\bibnamefont {Verma}}, \bibinfo {author} {\bibfnamefont {S.~W.}\
			\bibnamefont {Nam}}, \bibinfo {author} {\bibfnamefont {F.}~\bibnamefont
			{Marsili}}, \bibinfo {author} {\bibfnamefont {M.~D.}\ \bibnamefont {Shaw}},
		\bibinfo {author} {\bibfnamefont {A.~D.}\ \bibnamefont {Beyer}}, \ and\
		\bibinfo {author} {\bibfnamefont {A.}~\bibnamefont {Faraon}},\ }\href
	{http://science.sciencemag.org/content/early/2017/09/05/ science.aan5959}
	{\bibfield  {journal} {\bibinfo  {journal} {Science}\ }\textbf {\bibinfo
			{volume} {69}},\ \bibinfo {pages} {eaan5959} (\bibinfo {year}
		{2017})}\BibitemShut {NoStop}%
	\bibitem [{\citenamefont {Kis}\ \emph {et~al.}(2014)\citenamefont {Kis},
		\citenamefont {Mandula}, \citenamefont {Lengyel}, \citenamefont {Hajdara},
		\citenamefont {Kovacs},\ and\ \citenamefont {Imlau}}]{Kis:2014fq}%
	\BibitemOpen
	\bibfield  {author} {\bibinfo {author} {\bibfnamefont {Z.}~\bibnamefont
			{Kis}}, \bibinfo {author} {\bibfnamefont {G.}~\bibnamefont {Mandula}},
		\bibinfo {author} {\bibfnamefont {K.}~\bibnamefont {Lengyel}}, \bibinfo
		{author} {\bibfnamefont {I.}~\bibnamefont {Hajdara}}, \bibinfo {author}
		{\bibfnamefont {L.}~\bibnamefont {Kovacs}}, \ and\ \bibinfo {author}
		{\bibfnamefont {M.}~\bibnamefont {Imlau}},\ }\href
	{http://www.sciencedirect.com/science/article/pii/ S0925346714004273}
	{\bibfield  {journal} {\bibinfo  {journal} {Opt. Mat.}\ }\textbf {\bibinfo
			{volume} {37}},\ \bibinfo {pages} {845} (\bibinfo {year} {2014})}\BibitemShut
	{NoStop}%
	\bibitem [{\citenamefont {B{\"o}ttger}\ \emph {et~al.}(2016)\citenamefont
		{B{\"o}ttger}, \citenamefont {Thiel}, \citenamefont {Cone}, \citenamefont
		{Sun},\ and\ \citenamefont {Faraon}}]{Bottger:2016ix}%
	\BibitemOpen
	\bibfield  {author} {\bibinfo {author} {\bibfnamefont {T.}~\bibnamefont
			{B{\"o}ttger}}, \bibinfo {author} {\bibfnamefont {C.~W.}\ \bibnamefont
			{Thiel}}, \bibinfo {author} {\bibfnamefont {R.~L.}\ \bibnamefont {Cone}},
		\bibinfo {author} {\bibfnamefont {Y.}~\bibnamefont {Sun}}, \ and\ \bibinfo
		{author} {\bibfnamefont {A.}~\bibnamefont {Faraon}},\ }\href
	{https://journals.aps.org/prb/abstract/10.1103/ PhysRevB.94.045134}
	{\bibfield  {journal} {\bibinfo  {journal} {Phys. Rev. B}\ }\textbf {\bibinfo
			{volume} {94}},\ \bibinfo {pages} {045134} (\bibinfo {year}
		{2016})}\BibitemShut {NoStop}%
	\bibitem [{\citenamefont {Welinski}\ \emph {et~al.}(2016)\citenamefont
		{Welinski}, \citenamefont {Ferrier}, \citenamefont {Afzelius},\ and\
		\citenamefont {Goldner}}]{Welinski:2016gi}%
	\BibitemOpen
	\bibfield  {author} {\bibinfo {author} {\bibfnamefont {S.}~\bibnamefont
			{Welinski}}, \bibinfo {author} {\bibfnamefont {A.}~\bibnamefont {Ferrier}},
		\bibinfo {author} {\bibfnamefont {M.}~\bibnamefont {Afzelius}}, \ and\
		\bibinfo {author} {\bibfnamefont {P.}~\bibnamefont {Goldner}},\ }\href
	{\doibase 10.1103/PhysRevB.94.155116} {\bibfield  {journal} {\bibinfo
			{journal} {Phys. Rev. B}\ }\textbf {\bibinfo {volume} {94}},\ \bibinfo
		{pages} {155116} (\bibinfo {year} {2016})}\BibitemShut {NoStop}%
	\bibitem [{\citenamefont {Rakhmatullin}\ \emph {et~al.}(2009)\citenamefont
		{Rakhmatullin}, \citenamefont {Kurkin}, \citenamefont {Mamin}, \citenamefont
		{Orlinskii}, \citenamefont {Gafurov}, \citenamefont {Baibekov}, \citenamefont
		{Malkin}, \citenamefont {Gambarelli}, \citenamefont {Bertaina},\ and\
		\citenamefont {Barbara}}]{Rakhmatullin:2009jfa}%
	\BibitemOpen
	\bibfield  {author} {\bibinfo {author} {\bibfnamefont {R.~M.}\ \bibnamefont
			{Rakhmatullin}}, \bibinfo {author} {\bibfnamefont {I.~N.}\ \bibnamefont
			{Kurkin}}, \bibinfo {author} {\bibfnamefont {G.~V.}\ \bibnamefont {Mamin}},
		\bibinfo {author} {\bibfnamefont {S.~B.}\ \bibnamefont {Orlinskii}}, \bibinfo
		{author} {\bibfnamefont {M.~R.}\ \bibnamefont {Gafurov}}, \bibinfo {author}
		{\bibfnamefont {E.~I.}\ \bibnamefont {Baibekov}}, \bibinfo {author}
		{\bibfnamefont {B.~Z.}\ \bibnamefont {Malkin}}, \bibinfo {author}
		{\bibfnamefont {S.}~\bibnamefont {Gambarelli}}, \bibinfo {author}
		{\bibfnamefont {S.}~\bibnamefont {Bertaina}}, \ and\ \bibinfo {author}
		{\bibfnamefont {B.}~\bibnamefont {Barbara}},\ }\href {\doibase
		10.1103/PhysRevB.79.172408} {\bibfield  {journal} {\bibinfo  {journal} {Phys.
				Rev. B}\ }\textbf {\bibinfo {volume} {79}},\ \bibinfo {pages} {172408}
		(\bibinfo {year} {2009})}\BibitemShut {NoStop}%
	\bibitem [{\citenamefont {B{\"o}ttger}\ \emph {et~al.}(2006)\citenamefont
		{B{\"o}ttger}, \citenamefont {Thiel}, \citenamefont {Sun},\ and\
		\citenamefont {Cone}}]{Bottger:2006jo}%
	\BibitemOpen
	\bibfield  {author} {\bibinfo {author} {\bibfnamefont {T.}~\bibnamefont
			{B{\"o}ttger}}, \bibinfo {author} {\bibfnamefont {C.~W.}\ \bibnamefont
			{Thiel}}, \bibinfo {author} {\bibfnamefont {Y.}~\bibnamefont {Sun}}, \ and\
		\bibinfo {author} {\bibfnamefont {R.~L.}\ \bibnamefont {Cone}},\ }\href
	{\doibase 10.1103/PhysRevB.73.075101} {\bibfield  {journal} {\bibinfo
			{journal} {Phys. Rev. B}\ }\textbf {\bibinfo {volume} {73}},\ \bibinfo
		{pages} {075101} (\bibinfo {year} {2006})}\BibitemShut {NoStop}%
	\bibitem [{\citenamefont {Arcangeli}\ \emph {et~al.}(2014)\citenamefont
		{Arcangeli}, \citenamefont {Lovri{\'c}}, \citenamefont {Tumino},
		\citenamefont {Ferrier},\ and\ \citenamefont {Goldner}}]{Arcangeli:2014dr}%
	\BibitemOpen
	\bibfield  {author} {\bibinfo {author} {\bibfnamefont {A.}~\bibnamefont
			{Arcangeli}}, \bibinfo {author} {\bibfnamefont {M.}~\bibnamefont
			{Lovri{\'c}}}, \bibinfo {author} {\bibfnamefont {B.}~\bibnamefont {Tumino}},
		\bibinfo {author} {\bibfnamefont {A.}~\bibnamefont {Ferrier}}, \ and\
		\bibinfo {author} {\bibfnamefont {P.}~\bibnamefont {Goldner}},\ }\href
	{https://journals.aps.org/prb/abstract/10.1103/ PhysRevB.89.184305}
	{\bibfield  {journal} {\bibinfo  {journal} {Phys. Rev. B}\ }\textbf {\bibinfo
			{volume} {89}},\ \bibinfo {pages} {184305} (\bibinfo {year}
		{2014})}\BibitemShut {NoStop}%
	\bibitem [{\citenamefont {Guillot-No{\"e}l}\ \emph {et~al.}(2006)\citenamefont
		{Guillot-No{\"e}l}, \citenamefont {Goldner}, \citenamefont {Du},
		\citenamefont {Baldit}, \citenamefont {Monnier},\ and\ \citenamefont
		{Bencheikh}}]{2006PhRvB..74u4409G}%
	\BibitemOpen
	\bibfield  {author} {\bibinfo {author} {\bibfnamefont {O.}~\bibnamefont
			{Guillot-No{\"e}l}}, \bibinfo {author} {\bibfnamefont {P.}~\bibnamefont
			{Goldner}}, \bibinfo {author} {\bibfnamefont {Y.~L.}\ \bibnamefont {Du}},
		\bibinfo {author} {\bibfnamefont {E.}~\bibnamefont {Baldit}}, \bibinfo
		{author} {\bibfnamefont {P.}~\bibnamefont {Monnier}}, \ and\ \bibinfo
		{author} {\bibfnamefont {K.}~\bibnamefont {Bencheikh}},\ }\href {\doibase
		10.1103/PhysRevB.74.214409} {\bibfield  {journal} {\bibinfo  {journal}
			{Physical Review B - Condensed Matter and Materials Physics}\ }\textbf
		{\bibinfo {volume} {74}},\ \bibinfo {pages} {214409} (\bibinfo {year}
		{2006})}\BibitemShut {NoStop}%
	\bibitem [{\citenamefont {Fernandez-Gonzalvo}\ \emph
		{et~al.}(2015)\citenamefont {Fernandez-Gonzalvo}, \citenamefont {Chen},
		\citenamefont {Yin}, \citenamefont {Rogge},\ and\ \citenamefont
		{Longdell}}]{FG:2015dr}%
	\BibitemOpen
	\bibfield  {author} {\bibinfo {author} {\bibfnamefont {X.}~\bibnamefont
			{Fernandez-Gonzalvo}}, \bibinfo {author} {\bibfnamefont {Y.-H.}\ \bibnamefont
			{Chen}}, \bibinfo {author} {\bibfnamefont {C.}~\bibnamefont {Yin}}, \bibinfo
		{author} {\bibfnamefont {S.}~\bibnamefont {Rogge}}, \ and\ \bibinfo {author}
		{\bibfnamefont {J.~J.}\ \bibnamefont {Longdell}},\ }\href {\doibase
		10.1103/PhysRevA.92.062313} {\bibfield  {journal} {\bibinfo  {journal} {Phys.
				Rev. A}\ }\textbf {\bibinfo {volume} {92}},\ \bibinfo {pages} {062313}
		(\bibinfo {year} {2015})}\BibitemShut {NoStop}%
	\bibitem [{\citenamefont {Mims}(1968)}]{Mims:1968dg}%
	\BibitemOpen
	\bibfield  {author} {\bibinfo {author} {\bibfnamefont {W.~B.}\ \bibnamefont
			{Mims}},\ }\href {\doibase 10.1103/PhysRev.168.370} {\bibfield  {journal}
		{\bibinfo  {journal} {Phys. Rev.}\ }\textbf {\bibinfo {volume} {168}},\
		\bibinfo {pages} {370} (\bibinfo {year} {1968})}\BibitemShut {NoStop}%
	\bibitem [{\citenamefont {Viola}\ and\ \citenamefont
		{Lloyd}(1998)}]{Viola:1998ky}%
	\BibitemOpen
	\bibfield  {author} {\bibinfo {author} {\bibfnamefont {L.}~\bibnamefont
			{Viola}}\ and\ \bibinfo {author} {\bibfnamefont {S.}~\bibnamefont {Lloyd}},\
	}\href {\doibase 10.1103/PhysRevA.58.2733} {\bibfield  {journal} {\bibinfo
			{journal} {Phys. Rev. A}\ }\textbf {\bibinfo {volume} {58}},\ \bibinfo
		{pages} {2733} (\bibinfo {year} {1998})}\BibitemShut {NoStop}%
	\bibitem [{\citenamefont {Wen}\ \emph {et~al.}(2014)\citenamefont {Wen},
		\citenamefont {Duan}, \citenamefont {Ning}, \citenamefont {Huang},
		\citenamefont {Zhan}, \citenamefont {Zhang},\ and\ \citenamefont
		{Yin}}]{Wen:2014eh}%
	\BibitemOpen
	\bibfield  {author} {\bibinfo {author} {\bibfnamefont {J.}~\bibnamefont
			{Wen}}, \bibinfo {author} {\bibfnamefont {C.-K.}\ \bibnamefont {Duan}},
		\bibinfo {author} {\bibfnamefont {L.}~\bibnamefont {Ning}}, \bibinfo {author}
		{\bibfnamefont {Y.}~\bibnamefont {Huang}}, \bibinfo {author} {\bibfnamefont
			{S.}~\bibnamefont {Zhan}}, \bibinfo {author} {\bibfnamefont {J.}~\bibnamefont
			{Zhang}}, \ and\ \bibinfo {author} {\bibfnamefont {M.}~\bibnamefont {Yin}},\
	}\href {\doibase 10.1021/jp5050207} {\bibfield  {journal} {\bibinfo
			{journal} {J. Phys. Chem. A}\ }\textbf {\bibinfo {volume} {118}},\ \bibinfo
		{pages} {4988} (\bibinfo {year} {2014})}\BibitemShut {NoStop}%
	\bibitem [{\citenamefont {Kurkin}\ and\ \citenamefont
		{Chernov}(1980)}]{Kurkin:1980jx}%
	\BibitemOpen
	\bibfield  {author} {\bibinfo {author} {\bibfnamefont {I.~N.}\ \bibnamefont
			{Kurkin}}\ and\ \bibinfo {author} {\bibfnamefont {K.~P.}\ \bibnamefont
			{Chernov}},\ }\href {\doibase 10.1016/0378-4363(80)90107-2} {\bibfield
		{journal} {\bibinfo  {journal} {Physica B+C}\ }\textbf {\bibinfo {volume}
			{101}},\ \bibinfo {pages} {233} (\bibinfo {year} {1980})}\BibitemShut
	{NoStop}%
	\bibitem [{supp()}]{supp}%
	\BibitemOpen
	\bibinfo {note} {See Supplemental Material}\BibitemShut {NoStop}%
	\bibitem [{\citenamefont {Welinski}\ \emph {et~al.}(2017)\citenamefont
		{Welinski}, \citenamefont {Thiel}, \citenamefont {Dajczgewand}, \citenamefont
		{Ferrier}, \citenamefont {Cone}, \citenamefont {Macfarlane}, \citenamefont
		{Chaneliere}, \citenamefont {Louchet-Chauvet},\ and\ \citenamefont
		{Goldner}}]{Welinski:2017ew}%
	\BibitemOpen
	\bibfield  {author} {\bibinfo {author} {\bibfnamefont {S.}~\bibnamefont
			{Welinski}}, \bibinfo {author} {\bibfnamefont {C.~W.}\ \bibnamefont {Thiel}},
		\bibinfo {author} {\bibfnamefont {J.}~\bibnamefont {Dajczgewand}}, \bibinfo
		{author} {\bibfnamefont {A.}~\bibnamefont {Ferrier}}, \bibinfo {author}
		{\bibfnamefont {R.~L.}\ \bibnamefont {Cone}}, \bibinfo {author}
		{\bibfnamefont {R.~M.}\ \bibnamefont {Macfarlane}}, \bibinfo {author}
		{\bibfnamefont {T.}~\bibnamefont {Chaneliere}}, \bibinfo {author}
		{\bibfnamefont {A.}~\bibnamefont {Louchet-Chauvet}}, \ and\ \bibinfo {author}
		{\bibfnamefont {P.}~\bibnamefont {Goldner}},\ }\href {\doibase
		10.1016/j.optmat.2016.09.039} {\bibfield  {journal} {\bibinfo  {journal}
			{Opt. Mater.}\ }\textbf {\bibinfo {volume} {63}},\ \bibinfo {pages} {69}
		(\bibinfo {year} {2017})}\BibitemShut {NoStop}%
	\bibitem [{\citenamefont {Davies}(1974)}]{Davies:1974gu}%
	\BibitemOpen
	\bibfield  {author} {\bibinfo {author} {\bibfnamefont {E.~R.}\ \bibnamefont
			{Davies}},\ }\href {\doibase 10.1016/0375-9601(74)90078-4} {\bibfield
		{journal} {\bibinfo  {journal} {Phys. Lett. A}\ }\textbf {\bibinfo {volume}
			{47}},\ \bibinfo {pages} {1} (\bibinfo {year} {1974})}\BibitemShut {NoStop}%
	\bibitem [{\citenamefont {Schweiger}\ and\ \citenamefont
		{Jeschke}(2001)}]{schweiger2001principles}%
	\BibitemOpen
	\bibfield  {author} {\bibinfo {author} {\bibfnamefont {A.}~\bibnamefont
			{Schweiger}}\ and\ \bibinfo {author} {\bibfnamefont {G.}~\bibnamefont
			{Jeschke}},\ }\href@noop {} {\emph {\bibinfo {title} {{Principles of Pulse
					Electron Paramagnetic Resonance}}}}\ (\bibinfo  {publisher} {Oxford
		University Press},\ \bibinfo {year} {2001})\BibitemShut {NoStop}%
	\bibitem [{\citenamefont {George}\ \emph {et~al.}(2010)\citenamefont {George},
		\citenamefont {Witzel}, \citenamefont {Riemann}, \citenamefont {Abrosimov},
		\citenamefont {N\"otzel}, \citenamefont {Thewalt},\ and\ \citenamefont
		{Morton}}]{MOSiBiENDOR}%
	\BibitemOpen
	\bibfield  {author} {\bibinfo {author} {\bibfnamefont {R.~E.}\ \bibnamefont
			{George}}, \bibinfo {author} {\bibfnamefont {W.}~\bibnamefont {Witzel}},
		\bibinfo {author} {\bibfnamefont {H.}~\bibnamefont {Riemann}}, \bibinfo
		{author} {\bibfnamefont {N.~V.}\ \bibnamefont {Abrosimov}}, \bibinfo {author}
		{\bibfnamefont {N.}~\bibnamefont {N\"otzel}}, \bibinfo {author}
		{\bibfnamefont {M.~L.~W.}\ \bibnamefont {Thewalt}}, \ and\ \bibinfo {author}
		{\bibfnamefont {J.~J.~L.}\ \bibnamefont {Morton}},\ }\href {\doibase
		10.1103/PhysRevLett.105.067601} {\bibfield  {journal} {\bibinfo  {journal}
			{Phys. Rev. Lett.}\ }\textbf {\bibinfo {volume} {105}},\ \bibinfo {pages}
		{067601} (\bibinfo {year} {2010})}\BibitemShut {NoStop}%
	\bibitem [{\citenamefont {Tyryshkin}\ \emph {et~al.}(2006)\citenamefont
		{Tyryshkin}, \citenamefont {Morton}, \citenamefont {Ardavan},\ and\
		\citenamefont {Lyon}}]{Tyryshkin2006}%
	\BibitemOpen
	\bibfield  {author} {\bibinfo {author} {\bibfnamefont {A.~M.}\ \bibnamefont
			{Tyryshkin}}, \bibinfo {author} {\bibfnamefont {J.~J.~L.}\ \bibnamefont
			{Morton}}, \bibinfo {author} {\bibfnamefont {A.}~\bibnamefont {Ardavan}}, \
		and\ \bibinfo {author} {\bibfnamefont {S.~A.}\ \bibnamefont {Lyon}},\ }\href
	{\doibase 10.1063/1.2204915} {\bibfield  {journal} {\bibinfo  {journal} {J.
				Chem. Phys.}\ }\textbf {\bibinfo {volume} {124}},\ \bibinfo {pages} {234508}
		(\bibinfo {year} {2006})}\BibitemShut {NoStop}%
	\bibitem [{\citenamefont {Van~Vleck}(1940)}]{VanVleck:1940ji}%
	\BibitemOpen
	\bibfield  {author} {\bibinfo {author} {\bibfnamefont {J.~H.}\ \bibnamefont
			{Van~Vleck}},\ }\href {\doibase 10.1103/PhysRev.57.426} {\bibfield  {journal}
		{\bibinfo  {journal} {Phys. Rev.}\ }\textbf {\bibinfo {volume} {57}},\
		\bibinfo {pages} {426} (\bibinfo {year} {1940})}\BibitemShut {NoStop}%
	\bibitem [{\citenamefont {Shrivastava}(1983)}]{Shrivastava:1983tr}%
	\BibitemOpen
	\bibfield  {author} {\bibinfo {author} {\bibfnamefont {K.~N.}\ \bibnamefont
			{Shrivastava}},\ }\href@noop {} {\bibfield  {journal} {\bibinfo  {journal}
			{Phys. Stat. Sol. B}\ }\textbf {\bibinfo {volume} {117}},\ \bibinfo {pages}
		{437} (\bibinfo {year} {1983})}\BibitemShut {NoStop}%
	\bibitem [{\citenamefont {Abragam}\ and\ \citenamefont
		{Bleaney}(2012)}]{abragam2012electron}%
	\BibitemOpen
	\bibfield  {author} {\bibinfo {author} {\bibfnamefont {A.}~\bibnamefont
			{Abragam}}\ and\ \bibinfo {author} {\bibfnamefont {B.}~\bibnamefont
			{Bleaney}},\ }\href {https://books.google.co.uk/books?id=ASNoAgAAQBAJ} {\emph
		{\bibinfo {title} {{Electron Paramagnetic Resonance of Transition Ions}}}},\
	Oxford Classic Texts in the Physical Sciences\ (\bibinfo  {publisher} {OUP
		Oxford},\ \bibinfo {year} {2012})\BibitemShut {NoStop}%
	\bibitem [{\citenamefont {Morton}\ \emph {et~al.}(2008)\citenamefont {Morton},
		\citenamefont {Lees}, \citenamefont {Hoffman},\ and\ \citenamefont
		{Stoll}}]{MORTON2008315}%
	\BibitemOpen
	\bibfield  {author} {\bibinfo {author} {\bibfnamefont {J.~J.}\ \bibnamefont
			{Morton}}, \bibinfo {author} {\bibfnamefont {N.~S.}\ \bibnamefont {Lees}},
		\bibinfo {author} {\bibfnamefont {B.~M.}\ \bibnamefont {Hoffman}}, \ and\
		\bibinfo {author} {\bibfnamefont {S.}~\bibnamefont {Stoll}},\ }\href
	{\doibase http://dx.doi.org/10.1016/j.jmr.2008.01.006} {\bibfield  {journal}
		{\bibinfo  {journal} {J. Magn. Reson.}\ }\textbf {\bibinfo {volume} {191}},\
		\bibinfo {pages} {315 } (\bibinfo {year} {2008})}\BibitemShut {NoStop}%
	\bibitem [{\citenamefont {Tyryshkin}\ \emph {et~al.}(2012)\citenamefont
		{Tyryshkin}, \citenamefont {Tojo}, \citenamefont {Morton}, \citenamefont
		{Riemann}, \citenamefont {Abrosimov}, \citenamefont {Becker}, \citenamefont
		{Pohl}, \citenamefont {Schenkel}, \citenamefont {Thewalt}, \citenamefont
		{Itoh},\ and\ \citenamefont {Lyon}}]{Tyryshkin:2012fi}%
	\BibitemOpen
	\bibfield  {author} {\bibinfo {author} {\bibfnamefont {A.~M.}\ \bibnamefont
			{Tyryshkin}}, \bibinfo {author} {\bibfnamefont {S.}~\bibnamefont {Tojo}},
		\bibinfo {author} {\bibfnamefont {J.~J.~L.}\ \bibnamefont {Morton}}, \bibinfo
		{author} {\bibfnamefont {H.}~\bibnamefont {Riemann}}, \bibinfo {author}
		{\bibfnamefont {N.~V.}\ \bibnamefont {Abrosimov}}, \bibinfo {author}
		{\bibfnamefont {P.}~\bibnamefont {Becker}}, \bibinfo {author} {\bibfnamefont
			{H.-J.}\ \bibnamefont {Pohl}}, \bibinfo {author} {\bibfnamefont
			{T.}~\bibnamefont {Schenkel}}, \bibinfo {author} {\bibfnamefont {M.~L.~W.}\
			\bibnamefont {Thewalt}}, \bibinfo {author} {\bibfnamefont {K.~M.}\
			\bibnamefont {Itoh}}, \ and\ \bibinfo {author} {\bibfnamefont {S.~A.}\
			\bibnamefont {Lyon}},\ }\href {\doibase 10.1038/nmat3182} {\bibfield
		{journal} {\bibinfo  {journal} {Nat. Mater.}\ }\textbf {\bibinfo {volume}
			{11}},\ \bibinfo {pages} {143} (\bibinfo {year} {2012})}\BibitemShut
	{NoStop}%
	\bibitem [{\citenamefont {Orbach}(1961)}]{Orbach:1961cd}%
	\BibitemOpen
	\bibfield  {author} {\bibinfo {author} {\bibfnamefont {R.}~\bibnamefont
			{Orbach}},\ }\href {\doibase 10.1098/rspa.1961.0211} {\bibfield  {journal}
		{\bibinfo  {journal} {Proc. R. Soc. Lond. A}\ }\textbf {\bibinfo {volume}
			{264}},\ \bibinfo {pages} {458} (\bibinfo {year} {1961})}\BibitemShut
	{NoStop}%
	\bibitem [{\citenamefont {Sun}\ \emph {et~al.}(2008)\citenamefont {Sun},
		\citenamefont {B\"ottger}, \citenamefont {Thiel},\ and\ \citenamefont
		{Cone}}]{PhysRevB.77.085124}%
	\BibitemOpen
	\bibfield  {author} {\bibinfo {author} {\bibfnamefont {Y.}~\bibnamefont
			{Sun}}, \bibinfo {author} {\bibfnamefont {T.}~\bibnamefont {B\"ottger}},
		\bibinfo {author} {\bibfnamefont {C.~W.}\ \bibnamefont {Thiel}}, \ and\
		\bibinfo {author} {\bibfnamefont {R.~L.}\ \bibnamefont {Cone}},\ }\href
	{\doibase 10.1103/PhysRevB.77.085124} {\bibfield  {journal} {\bibinfo
			{journal} {Phys. Rev. B}\ }\textbf {\bibinfo {volume} {77}},\ \bibinfo
		{pages} {085124} (\bibinfo {year} {2008})}\BibitemShut {NoStop}%
	\bibitem [{\citenamefont {Mikkelson}\ and\ \citenamefont
		{Stapleton}(1965)}]{Mikkelson:1965kb}%
	\BibitemOpen
	\bibfield  {author} {\bibinfo {author} {\bibfnamefont {R.~C.}\ \bibnamefont
			{Mikkelson}}\ and\ \bibinfo {author} {\bibfnamefont {H.~J.}\ \bibnamefont
			{Stapleton}},\ }\href {\doibase 10.1103/PhysRev.140.A1968} {\bibfield
		{journal} {\bibinfo  {journal} {Physical Review}\ }\textbf {\bibinfo {volume}
			{140}},\ \bibinfo {pages} {A1968} (\bibinfo {year} {1965})}\BibitemShut
	{NoStop}%
	\bibitem [{\citenamefont {Campos}\ \emph {et~al.}(2004)\citenamefont {Campos},
		\citenamefont {Denoyer}, \citenamefont {Jandl}, \citenamefont {Viana},
		\citenamefont {Vivien}, \citenamefont {Loiseau},\ and\ \citenamefont
		{Ferrand}}]{Campos:2004cv}%
	\BibitemOpen
	\bibfield  {author} {\bibinfo {author} {\bibfnamefont {S.}~\bibnamefont
			{Campos}}, \bibinfo {author} {\bibfnamefont {A.}~\bibnamefont {Denoyer}},
		\bibinfo {author} {\bibfnamefont {S.}~\bibnamefont {Jandl}}, \bibinfo
		{author} {\bibfnamefont {B.}~\bibnamefont {Viana}}, \bibinfo {author}
		{\bibfnamefont {D.}~\bibnamefont {Vivien}}, \bibinfo {author} {\bibfnamefont
			{P.}~\bibnamefont {Loiseau}}, \ and\ \bibinfo {author} {\bibfnamefont
			{B.}~\bibnamefont {Ferrand}},\ }\href {\doibase 10.1088/0953-8984/16/25/015}
	{\bibfield  {journal} {\bibinfo  {journal} {J. Phys. Condens. Matter}\
		}\textbf {\bibinfo {volume} {16}},\ \bibinfo {pages} {4579} (\bibinfo {year}
		{2004})}\BibitemShut {NoStop}%
	\bibitem [{\citenamefont {Kiel}\ and\ \citenamefont
		{Mims}(1967)}]{1967PhRv161.386K}%
	\BibitemOpen
	\bibfield  {author} {\bibinfo {author} {\bibfnamefont {A.}~\bibnamefont
			{Kiel}}\ and\ \bibinfo {author} {\bibfnamefont {W.~B.}\ \bibnamefont
			{Mims}},\ }\href {\doibase 10.1103/PhysRev.161.386} {\bibfield  {journal}
		{\bibinfo  {journal} {Phys. Rev.}\ }\textbf {\bibinfo {volume} {161}},\
		\bibinfo {pages} {386} (\bibinfo {year} {1967})}\BibitemShut {NoStop}%
	\bibitem [{\citenamefont {Senyshyn}\ \emph {et~al.}(2004)\citenamefont
		{Senyshyn}, \citenamefont {Kraus}, \citenamefont {Mikhailik},\ and\
		\citenamefont {Yakovyna}}]{Senyshyn:2004hk}%
	\BibitemOpen
	\bibfield  {author} {\bibinfo {author} {\bibfnamefont {A.}~\bibnamefont
			{Senyshyn}}, \bibinfo {author} {\bibfnamefont {H.}~\bibnamefont {Kraus}},
		\bibinfo {author} {\bibfnamefont {V.~B.}\ \bibnamefont {Mikhailik}}, \ and\
		\bibinfo {author} {\bibfnamefont {V.}~\bibnamefont {Yakovyna}},\ }\href
	{\doibase 10.1103/PhysRevB.70.214306} {\bibfield  {journal} {\bibinfo
			{journal} {Phys. Rev. B}\ }\textbf {\bibinfo {volume} {70}},\ \bibinfo
		{pages} {2581} (\bibinfo {year} {2004})}\BibitemShut {NoStop}%
	\bibitem [{\citenamefont {Viola}\ and\ \citenamefont
		{Knill}(2003)}]{Viola:2003fc}%
	\BibitemOpen
	\bibfield  {author} {\bibinfo {author} {\bibfnamefont {L.}~\bibnamefont
			{Viola}}\ and\ \bibinfo {author} {\bibfnamefont {E.}~\bibnamefont {Knill}},\
	}\href {\doibase 10.1103/PhysRevLett.90.037901} {\bibfield  {journal}
		{\bibinfo  {journal} {Phys. Rev. Lett.}\ }\textbf {\bibinfo {volume} {90}},\
		\bibinfo {pages} {R040301} (\bibinfo {year} {2003})}\BibitemShut {NoStop}%
	\bibitem [{\citenamefont {Khodjasteh}\ and\ \citenamefont
		{Viola}(2009)}]{Khodjasteh:2009ff}%
	\BibitemOpen
	\bibfield  {author} {\bibinfo {author} {\bibfnamefont {K.}~\bibnamefont
			{Khodjasteh}}\ and\ \bibinfo {author} {\bibfnamefont {L.}~\bibnamefont
			{Viola}},\ }\href {\doibase 10.1103/PhysRevLett.102.080501} {\bibfield
		{journal} {\bibinfo  {journal} {Phys. Rev. Lett.}\ }\textbf {\bibinfo
			{volume} {102}},\ \bibinfo {pages} {181} (\bibinfo {year}
		{2009})}\BibitemShut {NoStop}%
	\bibitem [{\citenamefont {Hu}\ and\ \citenamefont
		{Hartmann}(1974)}]{Hu:1974kx}%
	\BibitemOpen
	\bibfield  {author} {\bibinfo {author} {\bibfnamefont {P.}~\bibnamefont
			{Hu}}\ and\ \bibinfo {author} {\bibfnamefont {S.~R.}\ \bibnamefont
			{Hartmann}},\ }\href {\doibase 10.1103/PhysRevB.9.1} {\bibfield  {journal}
		{\bibinfo  {journal} {Phys. Rev. B}\ }\textbf {\bibinfo {volume} {9}},\
		\bibinfo {pages} {1} (\bibinfo {year} {1974})}\BibitemShut {NoStop}%
	\bibitem [{\citenamefont {Maryasov}\ \emph {et~al.}(1982)\citenamefont
		{Maryasov}, \citenamefont {Dzuba},\ and\ \citenamefont
		{Salikhov}}]{Maryasov:1982cg}%
	\BibitemOpen
	\bibfield  {author} {\bibinfo {author} {\bibfnamefont {A.~G.}\ \bibnamefont
			{Maryasov}}, \bibinfo {author} {\bibfnamefont {S.~A.}\ \bibnamefont {Dzuba}},
		\ and\ \bibinfo {author} {\bibfnamefont {K.~M.}\ \bibnamefont {Salikhov}},\
	}\href {\doibase 10.1016/0022-2364(82)90007-5} {\bibfield  {journal}
		{\bibinfo  {journal} {J. Magn. Reson.}\ }\textbf {\bibinfo {volume} {50}},\
		\bibinfo {pages} {432} (\bibinfo {year} {1982})}\BibitemShut {NoStop}%
	\bibitem [{\citenamefont {Klauder}\ and\ \citenamefont
		{Anderson}(1962)}]{Klauder:1962it}%
	\BibitemOpen
	\bibfield  {author} {\bibinfo {author} {\bibfnamefont {J.~R.}\ \bibnamefont
			{Klauder}}\ and\ \bibinfo {author} {\bibfnamefont {P.~W.}\ \bibnamefont
			{Anderson}},\ }\href {\doibase 10.1103/PhysRev.125.912} {\bibfield  {journal}
		{\bibinfo  {journal} {Physical Review}\ }\textbf {\bibinfo {volume} {125}},\
		\bibinfo {pages} {912} (\bibinfo {year} {1962})}\BibitemShut {NoStop}%
	\bibitem [{\citenamefont {Boscaino}\ and\ \citenamefont
		{Gelardi}(1992)}]{Boscaino:1992ef}%
	\BibitemOpen
	\bibfield  {author} {\bibinfo {author} {\bibfnamefont {R.}~\bibnamefont
			{Boscaino}}\ and\ \bibinfo {author} {\bibfnamefont {F.~M.}\ \bibnamefont
			{Gelardi}},\ }\href {\doibase 10.1103/PhysRevB.46.14550} {\bibfield
		{journal} {\bibinfo  {journal} {Phys. Rev. B}\ }\textbf {\bibinfo {volume}
			{46}},\ \bibinfo {pages} {14550} (\bibinfo {year} {1992})}\BibitemShut
	{NoStop}%
	\bibitem [{\citenamefont {Agnello}\ \emph {et~al.}(2001)\citenamefont
		{Agnello}, \citenamefont {Boscaino}, \citenamefont {Cannas},\ and\
		\citenamefont {Gelardi}}]{Agnello:2001jr}%
	\BibitemOpen
	\bibfield  {author} {\bibinfo {author} {\bibfnamefont {S.}~\bibnamefont
			{Agnello}}, \bibinfo {author} {\bibfnamefont {R.}~\bibnamefont {Boscaino}},
		\bibinfo {author} {\bibfnamefont {M.}~\bibnamefont {Cannas}}, \ and\ \bibinfo
		{author} {\bibfnamefont {F.~M.}\ \bibnamefont {Gelardi}},\ }\href {\doibase
		10.1103/PhysRevB.64.174423} {\bibfield  {journal} {\bibinfo  {journal} {Phys.
				Rev. B}\ }\textbf {\bibinfo {volume} {64}},\ \bibinfo {pages} {174423}
		(\bibinfo {year} {2001})}\BibitemShut {NoStop}%
	\bibitem [{\citenamefont {Wolfowicz}\ \emph {et~al.}(2012)\citenamefont
		{Wolfowicz}, \citenamefont {Simmons}, \citenamefont {Tyryshkin},
		\citenamefont {George}, \citenamefont {Riemann}, \citenamefont {Abrosimov},
		\citenamefont {Becker}, \citenamefont {Pohl}, \citenamefont {Lyon},
		\citenamefont {Thewalt},\ and\ \citenamefont {Morton}}]{Wolfowicz:2012foa}%
	\BibitemOpen
	\bibfield  {author} {\bibinfo {author} {\bibfnamefont {G.}~\bibnamefont
			{Wolfowicz}}, \bibinfo {author} {\bibfnamefont {S.}~\bibnamefont {Simmons}},
		\bibinfo {author} {\bibfnamefont {A.~M.}\ \bibnamefont {Tyryshkin}}, \bibinfo
		{author} {\bibfnamefont {R.~E.}\ \bibnamefont {George}}, \bibinfo {author}
		{\bibfnamefont {H.}~\bibnamefont {Riemann}}, \bibinfo {author} {\bibfnamefont
			{N.~V.}\ \bibnamefont {Abrosimov}}, \bibinfo {author} {\bibfnamefont
			{P.}~\bibnamefont {Becker}}, \bibinfo {author} {\bibfnamefont {H.-J.}\
			\bibnamefont {Pohl}}, \bibinfo {author} {\bibfnamefont {S.~A.}\ \bibnamefont
			{Lyon}}, \bibinfo {author} {\bibfnamefont {M.~L.~W.}\ \bibnamefont
			{Thewalt}}, \ and\ \bibinfo {author} {\bibfnamefont {J.~J.~L.}\ \bibnamefont
			{Morton}},\ }\href {\doibase 10.1103/PhysRevB.86.245301} {\bibfield
		{journal} {\bibinfo  {journal} {arXiv.org}\ ,\ \bibinfo {pages} {245301}}
		(\bibinfo {year} {2012})}\BibitemShut {NoStop}%
	\bibitem [{\citenamefont {Souza}\ \emph {et~al.}(2012)\citenamefont {Souza},
		\citenamefont {Alvarez},\ and\ \citenamefont {Suter}}]{2012RSPTA.370.4748S}%
	\BibitemOpen
	\bibfield  {author} {\bibinfo {author} {\bibfnamefont {A.~M.}\ \bibnamefont
			{Souza}}, \bibinfo {author} {\bibfnamefont {G.~A.}\ \bibnamefont {Alvarez}},
		\ and\ \bibinfo {author} {\bibfnamefont {D.}~\bibnamefont {Suter}},\ }\href
	{\doibase 10.1098/rsta.2011.0355} {\bibfield  {journal} {\bibinfo  {journal}
			{Phil. Trans. R. Soc. A}\ }\textbf {\bibinfo {volume} {370}},\ \bibinfo
		{pages} {4748} (\bibinfo {year} {2012})}\BibitemShut {NoStop}%
	\bibitem [{\citenamefont {Viola}(2013)}]{Viola:book}%
	\BibitemOpen
	\bibfield  {author} {\bibinfo {author} {\bibfnamefont {L.}~\bibnamefont
			{Viola}},\ }\enquote {\bibinfo {title} {Quantum error correction},}\ \
	(\bibinfo  {publisher} {Cambridge University Press},\ \bibinfo {year}
	{2013})\ Chap.\ \bibinfo {chapter} {Introduction to quantum dynamical
		decoupling}\BibitemShut {NoStop}%
	\bibitem [{\citenamefont {Hahn}(1950)}]{PhysRev.80.580}%
	\BibitemOpen
	\bibfield  {author} {\bibinfo {author} {\bibfnamefont {E.~L.}\ \bibnamefont
			{Hahn}},\ }\href {\doibase 10.1103/PhysRev.80.580} {\bibfield  {journal}
		{\bibinfo  {journal} {Phys. Rev.}\ }\textbf {\bibinfo {volume} {80}},\
		\bibinfo {pages} {580} (\bibinfo {year} {1950})}\BibitemShut {NoStop}%
	\bibitem [{\citenamefont {Hu}\ and\ \citenamefont {Walker}(1978)}]{Hu:1978cj}%
	\BibitemOpen
	\bibfield  {author} {\bibinfo {author} {\bibfnamefont {P.}~\bibnamefont
			{Hu}}\ and\ \bibinfo {author} {\bibfnamefont {L.~R.}\ \bibnamefont
			{Walker}},\ }\href {\doibase 10.1103/PhysRevB.18.1300} {\bibfield  {journal}
		{\bibinfo  {journal} {Phys. Rev. B}\ }\textbf {\bibinfo {volume} {18}},\
		\bibinfo {pages} {1300} (\bibinfo {year} {1978})}\BibitemShut {NoStop}%
	\bibitem [{\citenamefont {Marino}(2011)}]{marino:pastel-00746050}%
	\BibitemOpen
	\bibfield  {author} {\bibinfo {author} {\bibfnamefont {R.}~\bibnamefont
			{Marino}},\ }\emph {\bibinfo {title} {{Proprietes magnetiques et optiques de
				cristaux dopes terres rares pour l'information quantique}}},\ \href
	{https://pastel.archives-ouvertes.fr/pastel-00746050} {\bibinfo {type}
		{Theses}},\ \bibinfo  {school} {{Universit{\'e} des Sciences et Technologie
			de Lille - Lille I}} (\bibinfo {year} {2011})\BibitemShut {NoStop}%
	\bibitem [{\citenamefont {Roberts}\ \emph {et~al.}(1999)\citenamefont
		{Roberts}, \citenamefont {Taylor}, \citenamefont {Gateva-Kostova},
		\citenamefont {Clarke}, \citenamefont {Rowley},\ and\ \citenamefont
		{Gill}}]{PhysRevA.60.2867}%
	\BibitemOpen
	\bibfield  {author} {\bibinfo {author} {\bibfnamefont {M.}~\bibnamefont
			{Roberts}}, \bibinfo {author} {\bibfnamefont {P.}~\bibnamefont {Taylor}},
		\bibinfo {author} {\bibfnamefont {S.~V.}\ \bibnamefont {Gateva-Kostova}},
		\bibinfo {author} {\bibfnamefont {R.~B.~M.}\ \bibnamefont {Clarke}}, \bibinfo
		{author} {\bibfnamefont {W.~R.~C.}\ \bibnamefont {Rowley}}, \ and\ \bibinfo
		{author} {\bibfnamefont {P.}~\bibnamefont {Gill}},\ }\href {\doibase
		10.1103/PhysRevA.60.2867} {\bibfield  {journal} {\bibinfo  {journal} {Phys.
				Rev. A}\ }\textbf {\bibinfo {volume} {60}},\ \bibinfo {pages} {2867}
		(\bibinfo {year} {1999})}\BibitemShut {NoStop}%
	\bibitem [{\citenamefont {Fauth}\ \emph {et~al.}(1986)\citenamefont {Fauth},
		\citenamefont {Schweiger}, \citenamefont {Braunschweiler}, \citenamefont
		{Forrer},\ and\ \citenamefont {Ernst}}]{Fauth:1986jt}%
	\BibitemOpen
	\bibfield  {author} {\bibinfo {author} {\bibfnamefont {J.~M.}\ \bibnamefont
			{Fauth}}, \bibinfo {author} {\bibfnamefont {A.}~\bibnamefont {Schweiger}},
		\bibinfo {author} {\bibfnamefont {L.}~\bibnamefont {Braunschweiler}},
		\bibinfo {author} {\bibfnamefont {J.}~\bibnamefont {Forrer}}, \ and\ \bibinfo
		{author} {\bibfnamefont {R.~R.}\ \bibnamefont {Ernst}},\ }\href {\doibase
		10.1016/0022-2364(86)90105-8} {\bibfield  {journal} {\bibinfo  {journal} {J.
				Magn. Reson.}\ }\textbf {\bibinfo {volume} {66}},\ \bibinfo {pages} {74}
		(\bibinfo {year} {1986})}\BibitemShut {NoStop}%
	\bibitem [{\citenamefont {Pla}\ \emph {et~al.}(2013)\citenamefont {Pla},
		\citenamefont {Tan}, \citenamefont {Dehollain}, \citenamefont {Lim},
		\citenamefont {Morton}, \citenamefont {Zwanenburg}, \citenamefont {Jamieson},
		\citenamefont {Dzurak},\ and\ \citenamefont {Morello}}]{Pla:2013io}%
	\BibitemOpen
	\bibfield  {author} {\bibinfo {author} {\bibfnamefont {J.~J.}\ \bibnamefont
			{Pla}}, \bibinfo {author} {\bibfnamefont {K.~Y.}\ \bibnamefont {Tan}},
		\bibinfo {author} {\bibfnamefont {J.~P.}\ \bibnamefont {Dehollain}}, \bibinfo
		{author} {\bibfnamefont {W.~H.}\ \bibnamefont {Lim}}, \bibinfo {author}
		{\bibfnamefont {J.~J.~L.}\ \bibnamefont {Morton}}, \bibinfo {author}
		{\bibfnamefont {F.~A.}\ \bibnamefont {Zwanenburg}}, \bibinfo {author}
		{\bibfnamefont {D.~N.}\ \bibnamefont {Jamieson}}, \bibinfo {author}
		{\bibfnamefont {A.~S.}\ \bibnamefont {Dzurak}}, \ and\ \bibinfo {author}
		{\bibfnamefont {A.}~\bibnamefont {Morello}},\ }\href {\doibase
		10.1038/nature12011} {\bibfield  {journal} {\bibinfo  {journal} {Nature}\
		}\textbf {\bibinfo {volume} {496}},\ \bibinfo {pages} {334} (\bibinfo {year}
		{2013})}\BibitemShut {NoStop}%
	\bibitem [{\citenamefont {Bushev}\ \emph {et~al.}(2011)\citenamefont {Bushev},
		\citenamefont {Feofanov}, \citenamefont {Rotzinger}, \citenamefont
		{Protopopov}, \citenamefont {Cole}, \citenamefont {Wilson}, \citenamefont
		{Fischer}, \citenamefont {Lukashenko},\ and\ \citenamefont
		{Ustinov}}]{Bushev:2011be}%
	\BibitemOpen
	\bibfield  {author} {\bibinfo {author} {\bibfnamefont {P.}~\bibnamefont
			{Bushev}}, \bibinfo {author} {\bibfnamefont {A.~K.}\ \bibnamefont
			{Feofanov}}, \bibinfo {author} {\bibfnamefont {H.}~\bibnamefont {Rotzinger}},
		\bibinfo {author} {\bibfnamefont {I.}~\bibnamefont {Protopopov}}, \bibinfo
		{author} {\bibfnamefont {J.~H.}\ \bibnamefont {Cole}}, \bibinfo {author}
		{\bibfnamefont {C.~M.}\ \bibnamefont {Wilson}}, \bibinfo {author}
		{\bibfnamefont {G.}~\bibnamefont {Fischer}}, \bibinfo {author} {\bibfnamefont
			{A.}~\bibnamefont {Lukashenko}}, \ and\ \bibinfo {author} {\bibfnamefont
			{A.~V.}\ \bibnamefont {Ustinov}},\ }\href {\doibase
		10.1103/PhysRevB.84.060501} {\bibfield  {journal} {\bibinfo  {journal}
			{Physical Review B}\ }\textbf {\bibinfo {volume} {84}},\ \bibinfo {pages}
		{060501} (\bibinfo {year} {2011})}\BibitemShut {NoStop}%
	\bibitem [{\citenamefont {Sukhanov}\ \emph {et~al.}(2017)\citenamefont
		{Sukhanov}, \citenamefont {Tarasov}, \citenamefont {Eremina}, \citenamefont
		{Yatsyk}, \citenamefont {Likerov}, \citenamefont {Shestakov}, \citenamefont
		{Zavartsev}, \citenamefont {Zagumennyi},\ and\ \citenamefont
		{Kutovoi}}]{Sukhanov:2017gx}%
	\BibitemOpen
	\bibfield  {author} {\bibinfo {author} {\bibfnamefont {A.~A.}\ \bibnamefont
			{Sukhanov}}, \bibinfo {author} {\bibfnamefont {V.~F.}\ \bibnamefont
			{Tarasov}}, \bibinfo {author} {\bibfnamefont {R.~M.}\ \bibnamefont
			{Eremina}}, \bibinfo {author} {\bibfnamefont {I.~V.}\ \bibnamefont {Yatsyk}},
		\bibinfo {author} {\bibfnamefont {R.~F.}\ \bibnamefont {Likerov}}, \bibinfo
		{author} {\bibfnamefont {A.~V.}\ \bibnamefont {Shestakov}}, \bibinfo {author}
		{\bibfnamefont {Y.~D.}\ \bibnamefont {Zavartsev}}, \bibinfo {author}
		{\bibfnamefont {A.~I.}\ \bibnamefont {Zagumennyi}}, \ and\ \bibinfo {author}
		{\bibfnamefont {S.~A.}\ \bibnamefont {Kutovoi}},\ }\href {\doibase
		10.1007/s00723-017-0888-7} {\bibfield  {journal} {\bibinfo  {journal} {Appl.
				Magn. Reson.}\ }\textbf {\bibinfo {volume} {131}},\ \bibinfo {pages} {1}
		(\bibinfo {year} {2017})}\BibitemShut {NoStop}%
\end{thebibliography}
\end{document}